\documentstyle[aas2pp4]{article}

\slugcomment{3/27/00 submitted ms.tex}

\lefthead{Zink, Lester, Doppmann,\& Harvey}
\righthead{Size of FIR Luminous Galaxies}

\begin{document}

\title{The Structure of IR Luminous Galaxies at 100 Microns}

\author{Eric C. Zink, Dan F. Lester, Greg Doppmann, Paul M. Harvey} 
\affil{Astronomy Department and McDonald Observatory, University of Texas,
    Austin, TX 78712}

\begin{abstract}
We have observed twenty two galaxies at 100 \micron\ with the Kuiper Airborne Observatory in order to determine the size of their FIR emitting regions. Most of these galaxies are luminous far-infrared sources, with $L_{FIR} > 10^{11} L_{\sun}$. This data constitutes the highest spatial resolution ever achieved on luminous galaxies in the far infrared. Our data includes direct measurements of the spatial structure of the sources, in which we look for departures from point source profiles. Additionally, comparison of our small beam 100 \micron\ fluxes with the large beam IRAS fluxes shows how much flux falls beyond our detectors but within the IRAS beam. Several sources with point-like cores show evidence for such a net flux deficit. We clearly resolved six of these galaxies at 100 \micron\ and have some evidence for extension in seven others. Those galaxies which we have resolved can have little of their 100 \micron\ flux directly emitted by a point-like active galactic nucleus (AGN). Dust heated to $\sim$40 K by recent bursts of non-nuclear star formation provides the best explanation for their extreme FIR luminosity. In a few cases, heating of an extended region by a compact central source is also a plausible option. Assuming the FIR emission we see is from dust, we also use the sizes we derive to find the dust temperatures and optical depths at 100 \micron\ which we translate into an effective visual extinction through the galaxy. Our work shows that studies of the far infrared structure of luminous infrared galaxies is clearly within the capabilities of new generation far infrared instrumentation, such as SOFIA and SIRTF.

\end{abstract}


\section{Introduction}

Data from the Kuiper Airborne Observatory, and later the Infra-Red Astronomical 
Satellite (IRAS) revealed that many galaxies emit much more flux in the far-
infrared (FIR, $\sim$40 - 120 \micron) than in any other wavelength band 
(\cite{Tel80}; \cite{Hou84}; \cite{Soi84}; \cite{deJ84}).  This FIR is usually attributed to thermal emission from warm ($\sim$40 K) dust heated by starbursts (\cite{Els85}; \cite{Law86}). There may also be a direct nonthermal contribution to the FIR luminosity from an active galactic nucleus (AGN), but at least the energy distributions suggest that emission from warm dust grains likely dominates the FIR flux even if there is an obvious active core and the galaxy is not obviously dusty (\cite{Bar92}; \cite{Chi92a}; \cite{Soi87}; \cite{Ede87a}).

While the most likely scenario for the FIR emission in galaxies is dust 
absorbing strong radiation from a burst of young massive stars, other scenarios 
are possible.  Extended dust could also be heated directly by an AGN core.  In 
addition, the dust could be reprocessing less energetic photons from 
an older population of stars in the galaxy, an energy source that is known to 
dominate in relatively low luminosity quiescent systems. It has been suggested 
that the dust could be heated by hot ($10^{8}$ K) intergalactic gas 
(\cite{Dwe92}; \cite{Bre90}), but this would only happen when such hot gas 
exists in the vicinity, as in the center of a large cluster of galaxies.  Dust could also be heated by shocks in the interstellar medium during the collision or interaction of galaxies (\cite{Har87}).

Each of the dust heating mechanisms above should correspond to particular 
spatial distributions of FIR light.  If the FIR emission is from the active core 
itself, the flux should appear point-like.  If it is from dust heated by an 
active nucleus, we might expect color temperature gradients to point toward the 
nucleus. Starburst heated dust should have about the same scale size as the 
burst itself because the young stars are well mixed with the gas from which the 
stars are forming and the optical depth for their UV photons is very high in the 
interstellar medium.  On the other hand, dust heated by nonionizing photons from 
a cooler population of stars that expel condensables might be expected to follow 
the smooth distribution of older stars in the galaxy. It is well understood that 
evidence for interactions, collisions, and mergers is almost invariably 
associated with the most luminous infrared galaxies, and comparison of the far 
infrared luminosity distribution with that of the interaction geometry is a 
vital clue to the mechanism by which this energy is produced. 

The question that motivated this observational project was a straightforward 
one. Does the far infrared emission in luminous galaxies, representing the bulk 
of the emitted energy, show any evidence for spatial structure using the highest 
spatial resolutions that are available to us? The distribution of luminosity in 
these sources can be a key to understanding their energy production mechanisms. 
Our observations address this question, and we develop relevant observational 
strategies for future high resolution studies. An associated question is whether 
we can use these measurements to help distinguish between the various plausible 
heating mechanisms.

High spatial resolution observations in the far infrared are seriously 
handicapped compared with observations in most other spectral regions. This part 
of the spectrum, containing the peak of energy emission from luminous galaxies, 
is inaccessible from the ground, and the spatial resolution that can be achieved 
is completely diffraction limited. A large aperture above the terrestrial water 
vapor absorption is a necessary tool. While large ground-based telescopes can 
attack the problem on arcsecond scales using tracers of hot and cool dust in the 
mid-IR and submillimeter continuum repectively, the correspondence of this dust 
with the distribution of warm dust that dominates the energy budget of luminous 
galaxies is not completely understood. To date, the largest telescopes that routinely observe in this spectral region are airborne and balloon-borne facilities.

While its high sensitivity allowed it to catalog many sources, IRAS resolved 
only the largest and nearest galaxies because of its comparatively large beam 
size of $\sim$2-4 arcmin at 60 and 100 \micron. Deconvolution efforts improved 
on this resolution somewhat, but could not approach the $\sim$23 arcsec 
diffraction-limited beam of the KAO. While ISO too has made fine contributions 
to the study of luminous galaxies, the higher S/N ratios that could be achieved 
in the far infrared continuum with ISOPHOT could not offset the substantially 
larger diffraction-limited beamsize of that telescope, for example the $\sim$43 
arcsec pixel size of the C100 channel.

We observed the spatial structure of a sample of luminous galaxies with the 
Kuiper Airborne Observatory at 100 \micron. These observations were among the 
last made during the long mission of this highly successful platform, which 
concluded in 1996.  With its 0.9m aperture, the KAO allowed the highest possible 
diffraction-limited resolution at this wavelength. This work is likely to be the 
highest spatial resolution study available near the peak of the energy emission 
from these galaxies until the commissioning of SOFIA and SIRTF. 

Determination of the distribution of far infrared emission in luminous galaxies 
enables several phenomenological and astrophysical insights into these objects. 
Many of these galaxies are interacting systems, and little direct information is 
available about precisely where in these multiple objects the luminosity 
actually originates. Such information would greatly illuminate the mechanism by 
which galactic collisions produce these luminous sources. The scale size over 
which the warm dust is distributed allows, through a simple, single-slab model 
calculation, the optical depth of the emitting dust. This information is of 
interest in our understanding of the global structure of these sources, and the 
extent to which higher spatial resolution information at shorter, more 
extinction-dependent wavelengths can be trusted to give an accurate picture. 
This analysis also allows independent estimation of the grain temperatures from 
the spectral color temperatures, which can be significantly different in sources 
with substantial optical depth.

In Section 2 we will discuss the sample of galaxies that we observed. Section 3 
will cover details about the observations and Section 4 will describe the 
analysis methodology. Section 5 presents the general results from our study, 
along with a discussion for each galaxy.  Further discussion of our results 
follows in Section 6.

\section{The Sample} \label{sample}

Our sample consisted of 22 galaxies with IRAS 100 \micron\ fluxes greater than 
or equal to 8 Jy.  Their targeting coordinates and the dates on which we 
observed them are listed in Table 1. We used the peak radio position of each 
galaxy for targeting because radio positions are more accurately known than the 
optical or FIR positions and are likely to correspond closely to the FIR peak 
(\cite{Bic90}). All of the galaxies except NGC 4151 (see also below) and UGC 
10923 are part of the Revised Bright Galaxy Sample (BGS) (\cite{BGS}) or 
the Bright Galaxy Survey - Part II (BGS2) (\cite{BGS2}).

The vagaries of airborne astronomy flight planning led to a somewhat 
inhomogeneous mix of objects. In general we attempted to observe the brightest 
sources possible and so we have a wide range of luminosities, but our 
observations were intentionally biased towards the highest. All of our sources 
except NGC 4151 and NGC 7625 have $L_{FIR} > 10^{10} L_{\sun}$ and 14 have a 
$L_{FIR} > 10^{11} L_{\sun}$. Physical characteristics of the sample galaxies 
are listed in Table 2. NGC 4151 is by far the least luminous galaxy we observed 
with a log$(L_{FIR}/L_{\sun}) = 9.5$ and was included in our sample 
specifically to confirm an earlier detection of extended emission (\cite{Eng88}; 
\cite{Gaf92}). As it turns out, it provides an excellent contrasting case to the others.

An additional selection constraint came from our attempt to select galaxies with 
relatively bright guide stars nearby. This was required because because the 
galaxy core was rarely bright enough to lock onto with the KAO optical guider. 
Because of this selection criterion, most of the galaxies are within 7\arcmin\ 
of an object with a magnitude V $<$ 12 as listed in the Hubble Telescope Guide 
Star Catalog (GSC). However the GSC does not list only stars, and several of 
the objects which we had planned to use as guide stars turned out to be brighter 
galaxies themselves.  So by preferentially choosing galaxies with nearby bright 
GSC objects, 16 of our 22 galaxies turned out to have obvious nearby companion 
galaxies.  The ratio of FIR luminous galaxies with companions to those without, 
in an unbiased sample, is between 1/4 and 1/8 (\cite{Soi84}) so even for our 
luminous galaxy sample, our ratio of $>$ 2/3 implies a significant bias towards 
galaxies with companions.

Other subtleties of airborne astronomy had a strong influence on the particular 
galaxies used in our sample but not their characteristics (aside from the bias 
noted above). The telescope in the airplane is essentially immobile in the 
azimuth direction so pointing in azimuth is accomplished by flying the airplane 
in the proper direction. The azimuth of the object being observed thus 
determines the direction of flight. The challenge of piecing together observing 
legs on relevant objects (such that one can eventually return to home base), and 
the need to observe the largest number of objects determined the time and 
azimuth at which each object was observed. Since the instrument was not equipped 
with an image rotator, the position angle of our linear array across each galaxy 
was not independently selectable, and the array remained oriented approximately 
in elevation. For sources in which a particular position angle was strongly 
preferred, this position angle became a flight planning constraint that 
sometimes could be achieved by observing at the right parallactic angle.

\section{Observations} \label{obs}

We observed all galaxies and calibration sources with the 0.9 m Kuiper Airborne Observatory (KAO) flying from Moffett Field, CA.  Data presented  in this paper is from flights on January 6, August 12, and August 16 in 1994 and on August 26 and 29 in 1995.  We used the University of Texas 2 $\times$ 10 $^{3}$He cooled silicon bolometer array with a filter centered near 100 \micron.  See \cite{Har79} for details of the ``100 \micron\ narrow" filter responsivity for this instrument. 

The detectors in the 2 $\times$ 10 spatial array are rectangular in shape with the short sides of the detectors aligned along the long axis of the array.  The long dimension of each detector corresponds approximately to the size of the diffraction spot of the KAO at 100 \micron\ ($\lambda$/D $\sim$ 23\arcsec).  The short dimension of each detector is half as large and thus critically samples the diffraction spot.  The detectors have center to center separation of 13.8\arcsec\ along the long axis of the array and 28\arcsec\ between the two arms (See Figure~\ref{fig1}) giving relatively little dead space between elements.

We observed in the nodding mode, alternating the position of the source between the two beams created by the $\sim$4\arcmin\ azimuth chop (along the short axis of the array).  In this mode a point source has a Gaussian FWHM of about 30\arcsec\ along the long axis of the array.  We scanned the detectors systematically over bright point sources during each flight series in order to determine individual detector responsivities for flat fielding.  These detector-to-detector responsivity ratios vary by about 5\% for different point sources on the same flight (\cite{Smi94}) and by a similar amount for point sources between flights during the same flight series.  This is far less than the photometric errors from other parts of our calibration (see Section~\ref{fluxcal}).

We targeted the telescope so that the peak of the FIR emission, which we presumed to correspond to the radio positions in Table 1, was on the center of the middle detector on one of the arms of the array (detector 5, see Figure~\ref{fig1} for definition of detector numbers and coordinates).  Only for NGC 4151 was the visual core bright enough that we could guide directly on it using the KAO focal plane video camera. 

The second arm of the array (detectors 11-20) was of lower quality, with several non-functional elements and others with high noise. As a result, we centered the source on the first arm (1-10) and used data from this second arm mostly as a check on positioning. The data from this second arm of the detector array is not presented here, though we show the locations of that second arm in the Figures to show the auxiliary information that was available to us. 

Since it was usually not possible to guide on the galaxy itself we guided on a nearby offset star.  On telescopes with an altitude-azimuth mount such as the KAO, the field of view rotates while tracking.  When offset guiding, this means the FIR source will appear to revolve around the guide star.  Only by knowing the field rotation can we accurately target the array at the FIR source. This angle is only known to $\sim$0.3\arcdeg\  on the KAO because of the limited resolution of the telescope stabilization system.  So the best pointing is achieved by using guide stars that are close to the galaxy, where the uncertainty in the field rotation will result in the smallest possible offset from the galaxy to the array center.  We consistently used guide stars $< 10$\arcmin\ away from the galaxy so the resulting tracking error was $< 3$\arcsec. This is a small fraction of the size of our pixels. The guide star positions were taken from the GSC, which can be assumed to be reliable at this level or better.

On the 16 August 1994 flight as well as the 1995 flights we either poorly defined the 100 \micron\ boresight on the array or the boresight drifted during the flight. The result of the poor boresight is that the center of the peaks shifted from the center of detector 5 by as much as 7\arcsec. The effects of the shift and how we dealt with them will be discussed in Section~\ref{fluxcal}.

\section{Data Reduction} \label{datared}
 
The images in Figures~\ref{fM+02} -~\ref{fU12915} show the detector array projected on the digitized Palomar Observatory Sky Survey (DSS) image with detector 5 centered at the position for each galaxy given in Table 1 (our target position), rotated to the average position angle at which we observed.  The range of the position angles over which the array moved during the time of observation is listed in Table 1, but for simplicity we do not show this on the figures. The change in position angles was usually fairly small, but in some cases was as much as 20\arcdeg\ over the course of our observation. Since the effect of this field rotation is to rotate the array around the target center, the outermost detectors (1 and 10) could have been, at times, as far as 12\arcsec\ from the positions shown in the Figures.  Usually the position angle changed by far less and in some cases we used the KAO ``freeze mode" which rotated the whole telescope at the same rate as the sky rotated so that the range in position angle was limited only by the telescope rotation stability (about 0.3\arcdeg). This was only possible for tracks in which the field rotated by less than the +/- 2 degree dynamic range of the telescope line-of-sight axis. 

\subsection{Flux Calibration} \label{fluxcal} 

Careful flux calibration might not seem necessary in view of the fact that we are mainly trying to determine the angular size and shape of the object. Nevertheless, good determinations of the emission of the galaxy in our relatively small beam combined with the larger beam observations from IRAS provide us with an additional test for spatial extension. If we get less signal in our peak detector than a point source with the cataloged IRAS flux should produce, we could interpret the deficit as evidence for extension on a scale of our detector size.

In order to make these comparisons worthwhile, we must make sure that our flux determinations are firmly on the IRAS photometric system. The IRAS 100 \micron\   primary standards were, however, all substantially resolved in our small beam, making them difficult to use as point-source comparisons. Therefore, we developed a secondary system of standards using asteroids, which are point-like for both instruments. The utility of such a system of standards has been recognized for ISOPHOT calibration (e.g. \cite{Mul98}).

In summary, this calibration effort involved using the radiometric constants derived from the IRAS Asteroid Survey (which give identically the IRAS fluxes for the relevant orbital geometry at the IRAS observation epoch), and used these radiometric constants to calculate their brightness at the particular epoch of our observations.  The asteroids are thus used as roving standards, where the thermophysical characteristics are taken as the calibration constants, and the predicted flux densities vary in a predictable way with the distances to the Sun and to the Earth.

Our choice of asteroids was constrained by those that were bright enough to give a high signal-to-noise detection, and those that were targetable in terms of the flight planning constraints detailed above. In addition, we rejected asteroids from our sample that were known to show more than ten percent optical variability which, whether because of albedo variations or non-sphericity might cause similar variations and predictive uncertainties at 100 \micron.

We calculated temperatures and fluxes of these asteroids with a simple thermal model (\cite{Leb86}).  The model assumes:  (1) spherical  shapes so that the flux does not vary with rotation, (2) slow rates of rotation, so that the dark side of the asteroid has time to cool down completely and only the sunlit side radiates in the IR, (3) albedos that are uniform and equal to those used in the IRAS Asteroid Survey, (4)  sizes from the IRAS asteroid survey, and (5) thermal equilibrium for all points on the asteroid.  We estimate an error of 10\% for the asteroid fluxes calculated with this model.  Deriving the system responsivities from the observed signal corrected for our filter response and the calculated asteroid fluxes puts us on the IRAS photometric system and allows comparisons between our fluxes and the fluxes from IRAS.

On 6 January 1994 we observed 1 Ceres, 52 Europa, and 97 Klotho in order to calibrate the galaxy fluxes.  We failed to observe an asteroid on 12 August 1994 and had to use the observation of 2 Pallas performed on 16 August to calibrate.  We believe the system responsivity was stable enough that this did not add significantly to the errors because we note the same signal strength on each night for those galaxies we observed on both nights.  We also found that for NGC 7541 which we only observed on 12 August, the flux we calculate matches the IRAS flux.  We calibrated with 704 Interamnia and 56 Melete on the 26 August 1995 flight, and with 704 Interamnia and 386 Siegena on the 29 August 1995 flight.

In addition to asteroids, we observed IRC+10420 as a point source calibrator and to define our boresight.  We cannot use IRC+10420 for photometry calibration because it seems to vary at 100 \micron\ from as low as $177 \pm 30$ Jy to as high as $360 \pm 108$ Jy (\cite{Har91}) and in the optical and near infrared as well (\cite{Jon93}). We also could not compare our observations of this object made on the flights of 12 and 16 August to make up for the lack of a calibrating asteroid observation on the 12 August flight.  A problem with the secondary chopping mirror which started after we observed IRC+10420 on 12 August and before we observed any galaxies changed the system sensitivity and prevents any simple comparison of IRC+10420 signals from the two nights.  The same chopping problem existed on the entire 16 August flight.

A one dimensional FIR profile was obtained for each galaxy and calibrator source by plotting the signal in each of the detectors 1-10. These plots are shown in Figures~\ref{fM+02} -~\ref{fU12915}. To these profiles we fit a baseline to the end detectors.  The baseline was usually the best fit line through detectors 1, 2, 9, and 10.  While the baseline allowed us to subtract small offsets that are the residual from chopping and nodding, that subtraction makes the derived fluxes entirely insensitive to emission on scales greater than $\sim$100\arcsec\ in the direction along the long axis of the array.  Optical images of each galaxy suggest that there is not likely to be a great deal of emission on scales larger than 100\arcsec\ scales unless the galaxy is very close. The small/large beam flux comparison provides a check on this assumption.

For each asteroid, we then added up the contributions from detectors 1-10, divided by the calculated flux of the asteroids, and divided the result by a factor 0.96 (which corrects our filter response to that of the IRAS bandpass) to get the system responsivity (signal per IRAS unit Jansky) at 100 \micron\ for detectors 1-10.  We did the same for the signal observed only in detector 5 to derive a measurement with a smaller effective aperture. See Table 3 for the actual signal levels (in DN) and responsivities for each of the asteroids.  For flights with more than one asteroid observation we use the average responsivity derived from all asteroids observed.  Our signal errors for the asteroids include not only errors due to noise from each of the detectors but also error associated with the uncertainty of the placement of the baseline.  These errors contribute approximately equally to the total error quoted.

Because of the stability of the system from flight to flight we might have expected the total system responsivities shown in the last 2 columns of Table 3 to be the same. The varying responsivities listed for the various flights differ because of changes in the system configuration between the flights and in particular the chopping problem experienced in 1995 August.

The profile of some asteroids and galaxies were not exactly centered in detector 5 of the detector array due to boresight shifts.  In order to compensate for such small pointing errors, we fit a Gaussian profile to each of the high S/N point source profiles of IRC+10420 obtained on the August flights and to the point-like 1 Ceres profile for the January 1994 flight.  We calculated the shift along the array needed to fit the point source Gaussian (with the same signal in detectors 1-10 as the galaxy or asteroid profile) to the two detectors in each source profile with the highest signal (usually detectors 4 and 5).  Since most sources had profiles that differed only slightly from that of a point source, the fit told us how far the center of the diffraction spot was from detector 5 along the long axis of the array as well as the signal that each detector should have had if the source had been perfectly centered.

We calculate the galaxy flux density by taking the signal from the galaxy, multiplying by the responsivity from the asteroids.  We then corrected for the response of the galaxy to the filter by dividing the result by a factor 0.87. This is a different factor than for the asteroids because of the much cooler spectral shape of the galaxies. Fluxes for the galaxies calculated from the signal in the central detector 5 only, corrected for any mispointing, are shown in the first column of Table 4. The detector 5 flux is simply the signal from that detector translated into the equivalent flux of a point source centered on detector 5.  The detector 5 flux should always be less than or equal to that summed across the whole array (detectors 1-10, also shown), but errors (mostly in baseline removal) allow some galaxies in Table 4 to have center detector fluxes slightly larger than their summed flux. As noted above, the errors for all fluxes include not only error associated with the individual detectors but also include some attempt to estimate the error associated with removing the baselines.  Error from the asteroid calibration (assumed to be 10\%) is also included in the errors quoted.  Our error estimates do not include small additional contributions from stratospheric opacity variations, filter corrections, pointing errors. It can be seen from the point source profiles in Figure 2 that the errors in offset guiding targeting discussed above should have only a minor effect on the signal that was measured. The effect of pointing errors on the measured source profile is discussed in more detail below.

In spite of the many steps and possible pitfalls in this calibration, it is gratifying that the measured galaxy fluxes never exceed the IRAS fluxes, which is consistent with a calibration of high accuracy.

Our observed small-beam fluxes are compared to the large-beam IRAS fluxes of these galaxies at 100 \micron\ in the BGS (\cite{BGS}) and the BGS2 (\cite{BGS2}).  For dust at 30-50 K with emissivity exponent n = 1-2, color correction changes the IRAS fluxes at 100 \micron\ by only about 5\%.  This is insignificant compared to the much larger errors ($\sim$15-20\%) of our calibration, so we do not correct the IRAS fluxes for color in Table 4 or anywhere in this paper. NGC 4151 and UGC 10923 are not in the BGS or the BGS2. The NGC 4151 flux is from Edelson, Malkan and Rieke (1987) and the UGC 10923 flux is from Mazzarella et al. 1991) In general there is excellent agreement between our observed and the (non-color corrected) IRAS fluxes.

\subsection{Resolution} \label{resolution} 

To determine if we resolved the galaxy or not, we compare the profile from detectors 1-10 with the point source profile (PSP).  We define the profile of IRC+10420 as the PSP on those flights where that bright source was observed.  For the 1994 January observations we used the profile of the high S/N asteroid 1 Ceres as our PSP. Note that while the fainter asteroids gave adequate S/N for flux calibration, IRC+10420 and Ceres profiles were most useful for generating accurate point source profiles.

If the galaxy is strongly emitting on scales greater than 20\arcsec, its profile will be noticeably more extended than the PSP.  In the upper parts of Figures~\ref{fM+02}-~\ref{fU12915} the profile of each galaxy we observed is compared to the PSP in two ways.  One PSP is scaled and shifted so that it is normalized to the peak signal of the galaxy (dashed lines), usually in detector 5.  The dotted lines are point sources with the IRAS flux shifted to the same offset as that of the Gaussian curve fit through the five central detectors in the profile. These latter dotted lines are what a point source with the IRAS 100  \micron\ flux would have looked like if it were centered on our array in the same way as the galaxy.

The width of the best fit Gaussian gives the first estimate of the size of the 
FIR emitting region.  If we assume the galaxy profile is intrinsically Gaussian then deconvolution is trivial, (FWHM$_{observed}^{2}$ = FWHM$_{source}^{2}$ 
+ FWHM$_{PSP}^{2}$), and allows us to determine the Gaussian size of the source, 
$D_{g}$ = FWHM$_{source}$.  In addition, we used a model of the galaxy disks 
such that $I = I_{o} exp(-r/r_{o}$) with a total flux equal to the flux we found 
from the sum of detectors 1-10.  We then found the best fit $r_{o}$ by a two 
dimensional convolution of the model with the two dimensional PSP (with a width 
of $\sim$40\arcsec\ in x).  The exponential disk size $D_{e}$ is then simply $2 
\times r_{o}$. The results of the fits to the profiles, $D_{g}$ and $D_{e}$   are given in Table 5.

If the profile does not show clear evidence for extended FIR emission by comparison with the shape of the PSP, then the photometry may reveal more subtle evidence for extension.  The IRAS beam is $\sim$2x4\arcmin\ at 100 \micron.  Our ``beam" for a single detector is $\sim$36\arcsec\ FWHM if we factor in the larger point source spread in the direction perpendicular to the long axis of the array.  Our ``beam" for the entire array is $\sim$40\arcsec\ $\times$ 100\arcsec.  If the flux we detect is smaller than the IRAS flux, and the source has not varied in brightness since the IRAS observations, we can argue that the spatial distribution of the FIR is extended such that some of the flux is outside of our beam but inside the IRAS beam.  Assuming the same exponential disk model mentioned above, we solved for $r_{o}$ for each galaxy using the detected flux from detector 5 and the IRAS flux.  These values are also included on Table 5. 

The sizes quoted in Table 5 can be considered significant only for those 
galaxies which show extension with respect to the point source profiles in 
Figures~\ref{fM+02}-~\ref{fU12915}.  Most of the galaxies with $D_{g} < 
20$\arcsec\ in Table 5 are indistinguishable from point sources and so the 
values given in Table 5 represent an upper limit to the size from our modeling.  
Those galaxies with $D_{g} > 20$\arcsec\ in Table 5 are convincingly extended 
with respect to the point source profiles, and the sizes quoted represent the 
actual size of the galaxy at 100 \micron.

The pointing errors mentioned in Sections~\ref{obs} and~\ref{fluxcal} would tend 
to make the galaxy profiles larger than they actually are because while we can 
directly guide on the optical images of the calibrating objects and assure that 
the position of the array does not change relative to them, we offset guide on 
most of the galaxies and thus have less assurance that the array did not shift 
significantly over the course of the observation.  Assuming we observed for half 
the time with the FIR core centered on detector 5 and half the time centered 
between detectors 4 and 5 ($\sim$7\arcsec\ away; matching the displacement of 
the calibrating asteroid observed at the end of the 16 August flight) the sizes 
we estimate in Table 5 for a galaxy with a $D_{g} \sim 20$\arcsec\ are only 10\% 
larger than they would be if the pointing were perfect.  The sizes for the 
galaxies with $D_{e} \sim 12$\arcsec\ are also only 10\% too large if we assume 
the diffraction spot moved with respect to the detector array. For galaxy FWHM 
smaller than $D_{g} \sim 20$\arcsec\ in the Gaussian model or $D_{e} \sim 
12$\arcsec\ in the exponential disk model, the errors due to possible 
mispointing are larger but we do not consider them resolved, in any case.

The widths of the point source profiles obtained on different flights and 
different flight series are the same to within 5\%. These errors are dominated 
by tracking errors. Assuming that the galaxy profiles have the same error in widths, standard propagation of errors means the values with Gaussian sizes of 
$\sim$20\arcsec\ given in Table 5 are good to within $\sim$15\%, not including 
the possible systematic error of up to 10\% due to mispointing.  For both the 
random error and the possible systematic error (due to pointing problems) the 
percentage of error in the quoted widths decreases with greater size, so those 
galaxies with Gaussian sizes much greater than 20\arcsec\ have more accurate 
sizes than those for the less extended galaxies.

\section{Results} \label{results} 

\subsection{Derived Characteristics} 

\subsubsection{Temperatures, Optical Depths, and Emissivity Exponents}

For a single uniform slab of dust, the flux observed at a particular wavelength 
$\lambda$ is \begin{equation} \label{eq1} S_{\lambda} 
=B_{\lambda}(T_{d})Q_{\lambda}\Omega_{\lambda}
\end{equation} where $B_{\lambda}(T_{d})$ is the Planck function at the dust 
temperature $T_{d}$, $Q_{\lambda}$ is the dust emissivity at $\lambda$, and 
$\Omega_{\lambda}$ is the apparent size of the slab.  Equation~\ref{eq1} is 
often used to derive a ``dust temperature" from the 60 and 
100 \micron\ IRAS FIR fluxes by assuming (1) the galaxy contains only a single 
slab of dust with a single temperature, (2) $\tau_{\lambda}$ is very small at 
FIR wavelengths so that $Q_{\lambda} \propto [1 - e^{-\tau_{\lambda}}]$ reduces 
to $Q_{\lambda} \propto \tau_{\lambda}$, and (3) $\tau_{\lambda} \propto 
\lambda^{-n}$ where n is the emissivity usually given a value 1-2.  Condition 
(1) above implies among other things that $\Omega_{60} = \Omega_{100}$ and 
conditions (2) and (3) give $Q_{60}/Q_{100} = (60/100)^{-n}$ so that with 
$S_{60}/S_{100}$, one can solve for $T_{d}$.

Clearly this method has its shortcomings.  Dust does not congregate in simple 
slabs at single temperatures and the temperature one calculates depends on the 
assumed value for.  In addition, colder dust emitting primarily at 
wavelengths longer than 100 \micron\ will not be detected.  Thus the temperature 
given by this single slab method cannot even be realistically classified as an 
average or median dust temperature but perhaps only as a typical temperature for 
the warm dust in the system.

Despite the shortcomings of the simple slab method, we use it to estimate dust 
temperature retaining assumptions (1) and (3).  The typically 2-3 points of FIR 
data available for each galaxy are not enough to support models of greater 
complexity, such as models with several slabs at different temperatures. 
Infrared Space Observatory (ISO) broad band observations of NGC 6090  
(\cite{Aco96}) show that both hotter and colder dust exists.  However, the 60 
and 100 \micron\ points still accurately model the dominant 20-50 K dust feature 
in both Seyfert (\cite{Rod96}) and starburst galaxies.  

In this analysis, we use the IRAS fluxes at both 60 and 100 \micron\ because we 
did not measure at 60 \micron\ and the large beam IRAS fluxes are more likely to 
be spatially consistent with each other. As will be discussed below, 
occasionally our 100 \micron\ flux is significantly different from the IRAS 
flux, usually because our smaller beam is pointed at only one of a cluster of 
galaxies which were all in the IRAS beam. In these cases we use our 100 \micron\ 
flux with a 60 \micron\ flux scaled so that our 60/100 ratio matches that of the 
IRAS fluxes. In this way, the $T_{d}$ we generate is always close to the $T_{d}$ 
that the IRAS fluxes would give without any further information.  It is also 
important to note that by using data from both our 100 \micron\ observations and 
IRAS fluxes from an epoch about 10 years earlier, we are assuming that the FIR 
fluxes of the galaxies we observed were constant from the IRAS epoch to that of 
our observation.

Our high spatial resolution data add the advantage of defining $\Omega_{100}$ 
and, by assumption (1) above, $\Omega_{\lambda}$ for the other wavelengths for 
which we have data. The spatial information allows us to calculate $\tau_{100}$ 
at the same time we as we derive $T_{d}$, but with a few notable exceptions of 
high optical depth, the $T_{d}$ as calculated by our simple model is insensitive 
to the dust region size. 

For Table 6, we define the angular size of a dust emission region in a galaxy, 
$\Omega_{100}$, as the size of a circle with the diameter of the galaxy size as 
defined or limited by $D_{e}$.  We use the scale length fit to the profile 
instead of the scale length derived from the flux in detector 5 and the IRAS 
flux because it not only more accurately measures the size of the 
core (and only the core) but also uses the information from many detectors 
instead of just one.  For Table 7 we use $D_{g}$ for those galaxies which we 
were able to resolve.  We assume an emissivity exponent n = 1.5 for Tables 6 and 
7.

For those galaxies with measurements at longer wavelengths in the literature, we  use the extra fluxes to derive n along with the derived dust temperature and the optical depth at 100 \micron.  The results are in Table 8; note that the value of n varies from 1.5 to greater than 2.2, but most estimates remain close to a typical galactic value of 
of n = 1.5. 

Having derived the optical depth, and assuming the dust is similar to dust in the solar neighborhood, we can use observed dust properties to estimate the visual extinction $A_{V}$.  We assume a value of $\sim$750 for $A_{V} / \tau_{100}$ (\cite{Mak85}). This extinction may be particularly important when we attempt, for example, to use the sizes of H$\alpha$ regions to estimate the size of the central starburst.  A high $A_{V}$ value implies that the H$\alpha$ size may be unreliable.  The calculated values of $A_{V}$ we used and what they imply for each galaxy are included in Section~\ref{discogal}.  They were taken from (in order of preference) Tables 8, 6, and 7.

It is important to note that the single slab method we use to determine $A_{V}$ 
is incapable of dealing with more complicated material distributions, and so it 
is risky to assume that our derived $A_{V}$ applies uniformly over the central emitting region of a galaxy.  It is much safer to use it to roughly estimate the amount of visual extinction from a FIR emitting region.

\subsubsection{The q-value} \label{qval}

The high resolution observations allow us to determine more accurately how much 
FIR flux is coming from galaxies and galaxy systems that were unresolved by 
IRAS.  For galaxy pairs this allows a determination of the q-value (the ratio of 
FIR to radio flux) for each galaxy in the pair instead of for the system as a 
whole.  For isolated galaxies it may allow the determination of this q-value for 
the central regions of the galaxy, independent of its outer regions.  

The q-value is of interest because it is very nearly constant for a wide variety 
of galaxies with luminosities between $10^{9}$ and $10^{13} L_{\sun}$ (\cite{Con91a}, hereafter CHYT). This constancy is remarkable because 
of the completely different mechanisms by which the FIR and radio emission is 
produced. One way to reconcile this is through star formation where new stars 
heat dust to cause the FIR radiation and produce supernovae which generate the 
nonthermal radio emission (\cite{Con91a}), though the time scales of the energy 
production is quite different in these two cases.

The q-value is defined (\cite{Hel85}) by \begin{equation} \label{eq2}
q \equiv \log [(FIR / 3.75 \times 10^{12} \mathrm{Hz}) / S_{1.4 \mathrm{GHz}}]
\end{equation} where $S_{1.4 \mathrm{GHz}}$ is in W m$^{-2}$ Hz$^{-1}$ and
\begin{equation} \label{eq3} FIR \equiv 1.26 \times 10^{-14} (2.58 S_{60} + 
S_{100}) \end{equation} is the estimated flux from 42.5 - 122.5 \micron\ (in W 
$m^{-2}$) if $S_{60}$ and $S_{100}$ are IRAS survey measurements in Jy 
uncorrected for color (\cite{Hel88}). These are listed in Table 2, and 
discussed individually in the sections below in which the new information about the far infrared morphology bears on it.

The better determination of q for an individual galaxy in a pair could help 
explain the source of an anomalous q-value for that galaxy pair.  For example in 
UGC 12914/5 (see Section~\ref{U12915} a ``radio bridge" between the pair lowers 
the system q-value, but we show the q for the FIR dominant UGC 12915 is normal.    
 
Our new calculations of the q-value might also help determine if the most 
luminous IR galaxies, and OH megamasers galaxies in particular, have an IR 
excess that gives a higher q-value. (\cite{Mar89}).  We observed four OH 
megamaser galaxies and, when using the IRAS fluxes, the q-value is too high in three of them (III Zw 35, IRAS 1720-00, and Zw 475.056).  Using our fluxes however, the q-value is too high in III Zw 35 and Zw 475.056 and in fact is low for UGC 08696.  Of course, a sample of four does not decide the issue, but we find little evidence for high q-values in OH megamaser galaxies. 

\subsection{Discussion of Individual Galaxies} \label{discogal}

\subsubsection{MCG+02-04-025} \label{M+02}

MCG+02-04-025 (IRAS 01173+1405) is a spiral with an optical extension of about 
18\arcsec\ $\times$ 30\arcsec\ (\cite{Mir88}).  In the DSS image there is a hint 
of a bridge of material between the galaxy and a fainter galaxy only 
$\sim$1\arcmin\ to the east (Figure~\ref{fM+02}). The interaction between the 
galaxies might be responsible for the log$(L_{FIR}/L_{\sun}) = 11.27$. 
Mirabel \& Sanders (1988) also note that the \ion{H}{1} line profile is not the 
``horned" double peak of a normal spiral but is a single peak more common to 
galaxies that have suffered some interaction. They detected MCG+02-04-025 in 
\ion{H}{1} in both emission and absorption.  From their detection they deduce an 
atomic gas mass greater than $3.4 \times 10^{9} M_{\sun}$, most of which must be 
concentrated near the core in order to provide a sufficient optical depth to 
explain the absorption. Our observations show this galaxy to be marginally 
resolved at 100 \micron\ (Figure~\ref{fM+02}), with a $D_{g} = 19.5$\arcsec\ and 
a $D_{e} = 11$\arcsec. The galaxy profile is slightly larger than the PSP, but 
better evidence comes from the photometry.  We do not recover the entire IRAS 
flux using the sum of detectors 1-10, and our detectors 11-20 show little 
evidence for any emission.  This leaves the other side of the detector 1-10 arm 
of the array (to the southeast) as the place that may have the missing IRAS 
flux.

Using the flux from the center detector 5, the exponential disk model fit to our 
data gives an $r_{o} \sim 6.5$\arcsec\ when compared to the larger IRAS beam 
measurement. This is close to but larger than the $r_{o} \sim 5$\arcsec\ derived 
using the same exponential disk model with multibeam photometry at K 
(\cite{Car90b}).  Using the size of 2x$r_{o} = 11$\arcsec\ , our 100 \micron\ 
flux, and a 60 \micron\ flux scaled so that the $S_{60}/S_{100}$ ratio is the 
same as the \cite{BGS} fluxes, we derive a $T_{d} = 44.6$ and $\tau_{100} = 2.1 
\times 10^{-3}$ from which we get an $A_{V} \sim 1.6$.  The $A_{V}$ we derive 
from the FIR fluxes and size is low considering the large gas column density 
implied by Mirabel \& Sanders (1988).

The possible extension at FIR wavelengths is interesting when compared to the 
radio morphology.  The estimated size of the emission region depends on the size 
of the interferometric array.  With a synthesized 6\arcsec\ beam (1.49 GHz, 
\cite{CHSS}) find a Gaussian size 2\arcsec\ $\times$ 2\arcsec.  With a smaller 
0.25\arcsec\ beam (8.44 GHz, CHYT) the size is only 1.2\arcsec\ $\times$ 0.8\arcsec.  Also, VLBI studies found an extremely compact core with emission on 5, 10, and 50 milliarcsecond scales (\cite{Lon93}; \cite{Smi98})

The radio studies argue quite convincingly that there is a compact radio core 
which might even be small enough that it becomes difficult to explain the 
luminosity with a starburst and its supernovae (\cite{Smi98}).  However, the 
fact that in each case the radio emission was resolved, along with the possible 
extension in the FIR, argue that while a central luminous radio core may exist 
(weak AGN or starburst), much of the FIR flux originates on larger spatial 
scales. If one assumes that the FIR flux is due to dust heated by a central 
compact source, a simple spherically symmetric dust model (\cite{Bar87}) 
suggests that this dust might have a size $\sim$19\arcsec. This size assumes 
that the luminosity of the central source in the UV matches the FIR luminosity 
and the dust at $T_{d} = 45$ has a clear view of the core. Thus, for this 
object, we conclude that while the far infrared emission is likely to be emitted 
in distributed sources, but that the situation energetically allows those 
sources to be heated by a central source, whether AGN or compact starburst.

\subsubsection{III Zw 35} \label{3Z35} 

III Zw 35 is a close pair in visible light with components to the northeast and 
southwest separated by about 10\arcsec\ (Figure~\ref{f3Z35}).  The brighter 
northern component contains an active core (Seyfert 2), an OH megamaser 
(\cite{Dia99}; \cite{Tro97}), and is also the likely source for the formaldehyde 
($H_{2}CO$) maser emission seen from the system (\cite{Baa93}).

We do not resolve III Zw 35 at 100 \micron\ (Figure~\ref{f3Z35}) and our 
observed flux is almost identical to the IRAS flux.  Using the BGS fluxes 
and the upper limit of the diameter of the emitting region of 12\arcsec, we 
obtain a dust temperature of $\sim$40 K and an optical depth at 100 \micron\ 
$\tau_{100} > 4.8 \times 10^{-3}$.  The implied $A_{V}$ is $>$ 3.6.

The galaxy is unresolved in \cite{CHSS} with a 1.49 GHz radio flux of 41.2 mJy 
but \cite{CHYT} resolve the galaxy at 8.44 GHz with a size of  0.18\arcsec\ 
$\times$ 0.14\arcsec.  The 1.49 GHz continuum flux with the FIR flux gives a 
somewhat high q-value of 2.56 which may be related to the OH  megamaser (see the 
discussion in Section~\ref{qval}).

\subsubsection{UGC 02369} \label{U2369} 

UGC 02369 is a close pair of galaxies aligned north-south and separated by about 
30\arcsec\ (Figure~\ref{fU2369}).  The northern component is brighter in the 
visible and NIR (\cite{Car90b}) but the radio positions and the fact that we 
recover essentially all of the IRAS flux (within our errors) while centered on 
the southern component establish that the southern galaxy produces almost all 
the FIR flux. Mid-infrared ISOCAM images (Hwang et al. 1999) also show this 
component to be dominant at those wavelengths. While VLBI maps of the source 
show some evidence for small scale structure, the emission appears to be too 
strong and compact to come from star forming regions. An AGN power source for 
the radio emission seems likely (\cite{Smi98}). Mirabel \& Sanders (1988) find 
\ion{H}{1} in absorption in UGC 02369 and estimate a column density of $2.2 
\times 10^{19}$ cm$^{-2}$ with an assumed spin temperature of 100 K.

We fail to resolve UGC 02369 either directly (Figure~\ref{fU2369}) or through 
our flux estimates, which are consistent with IRAS within the errors. Hwang et al. (1999) claim that the ISOCAM source is slightly resolved at 15 \micron\ in 
3\arcsec\ pixels. They do not cite a size explicitly, but it appears to be close 
to that of our 100 \micron\ size limit of 12\arcsec. Using this limit and the 
BGS 60 and 100 \micron\ fluxes we obtain a dust temperature of $\sim$36.6 
K and a visual extinction of greater than $\sim$3.9 for the system.  When 
combined with the CHSS 1.49 GHz radio flux of 50 mJy, the q-value is a 
normal 2.32.

\subsubsection{NGC 1275} \label{N1275} 

NGC 1275 is a giant elliptical galaxy (Figure~\ref{fN1275}) associated with the 
radio source 3C 84 at the core of the Perseus cluster.  Not only does this 
galaxy have a strong AGN but it also is associated with a cooling flow and two 
systems of low ionization filaments, one of which is probably the remnants of a 
recent merger.  See Lester et al. (1995) 
analysis of the 100 \micron\ flux from NGC 1275 with the same data.

The optical size from the Uppsala General Catalog of Galaxies (UGC, 
\cite{Nil73}) is 3.5\arcmin\ $\times$ 2.5\arcmin.  We clearly resolve the core 
at 100 \micron\ (Figure~\ref{fN1275}b) and show that $D_{g} = 30$\arcsec\ and 
$D_{e} = 17$\arcsec.  The exponential disk fit from fluxes gives a size of only 
$D_{e} \sim 13$\arcsec\ but given the uncertainty of the fluxes and the 
effective beam size, the 17\arcsec\ estimate is probably more accurate.  Using 
the single slab  model, the 60 and 100 \micron\ fluxes from the BGS2, and 
the diameter of the emitting region of 17\arcsec\ we obtain a dust temperature   
$T_{d}$ of 44 K and an optical depth of $8.4 \times 10^{-4}$ which corresponds 
to an $A_{V} \sim 0.6$.  Because NGC 1275 is not a spiral and the exponential 
disk model may not represent the galaxy well, we also note here that with the 
Gaussian $D_{g} =30$\arcsec\ the temperature remains nearly the same but the 
optical depth and corresponding visual extinction drop so that $A_{V} \sim 0.2$ 
(Table 7).  In any case $A_{V}$ is low enough that optical images should 
represent the population from the galaxy fairly well.

\subsubsection{VII Zw 31} \label{7Z31} 

This galaxy is only a fuzzy blob on the POSS (Figure~\ref{f7Z31}) with an 
optical extent of 10\arcsec\ or less.  According to optical profiles from
Djorgovski, de Carvalho, \& Thompson (1990) the FWHM is less than 5\arcsec. 
These authors also mention an object $\sim$20\arcsec\ to the northwest of the 
galaxy that is either a companion or the remains of a merger. The large 
luminosity in the FIR log$(L_{FIR}/L_{\sun}) = 11.6$ of VII Zw 31 only became 
apparent with the advent of IRAS (Fairclough, 1986) but its distance of 
$\sim$220 Mpc (Sanders, Scoville, \& Soifer 1991) suggests not only that its 
angular extent will be small but that its FIR flux will be small as well. 

Sage \& Solomon (1987) measured the CO emission from VII Zw 31 and discovered 
that  the galaxy contains $5 \times 10^{10} M_{\sun}$ of gas, which is roughly 
half of the dynamical mass of the galaxy.  Other observations (\cite{San91}; 
\cite{Rad91}; \cite{Sco89}) have confirmed this tremendous gas mass and show 
that the size of the CO emitting region is not resolved with a beamsize of 
7\arcsec.  The surface density of the gas is therefore must be $>
1000 M_{\sun}/\mathrm{pc}^{2}$.  This is $\sim$5 times greater than the mean 
surface density of $\sim 170 M_{\sun}/\mathrm{pc}^{2}$ of giant molecular clouds 
in our galaxy (\cite{Sag87}; \cite{Sol87}).  Because of the large gas mass,
Sage \& Solomon suggested that VII Zw 31 could be a proto-galactic disk that has 
yet to form most of its mass into stars.  However, Djorgovski et al. (1990) 
argue that VII Zw 31 is more likely a ``merger-induced starburst" that has yet 
to use up most of the gas acquired in the merger. New CO interferometer 
observations by Downes and Solomon (1998) clearly show evidence for a rapidly 
rotating nuclear ring on a scale of several hundred parsecs (a few arcseconds).

We did not resolve VII Zw 31 by its profile and our determination of its flux 
only marginally suggests spatial extension (Figure~\ref{f7Z31}). All of the FIR 
emission is thus likely to be produced within or around the CO emitting region. 
In our data of VII Zw 31 detector 1 gave an anomalous and inconsistent high 
flux.  Because of this, we took our baseline through detectors 2, 9, and 10, and 
our ``sum of detectors" flux is for detectors 2-10 only. Using the maximum size 
of the CO emitting region as the maximum extent of the FIR (Diameter = 
7\arcsec), the thermal dust model gives a dust temperature $\sim$34 K and an 
optical depth of $\tau_{100} > 2.0 \times 10^{-2}$. which corresponds to a large 
$A_{V} >$ 15.

\subsubsection{UGC 05101} \label{U5101} 

UGC 05101 is a disturbed spiral galaxy extended east-west in the optical with 
both a large ring (\cite{San88}; \cite{Whi90}), and a jet (\cite{San88}) or 
tidal tail (\cite{Maj93}) extending at least 40\arcsec\ to the west. This latter 
feature is just visible on the DSS images (Figure~\ref{fU5101}). There is only 
one core in near infrared images (\cite{Car90a}, \cite{Gen98}) and several 
theories have been advanced to account for the morphology, the FIR luminosity of 
log$(L_{FIR}/L_{\sun}) = 11.77$, and the presence of an active core with a 
LINER/Seyfert 2 emission spectrum (\cite{Maj93}). Sanders et al. (1988) 
explain the ring and active core of UGC 05101 as the result of an 
interaction with another gas rich spiral which they claim could have already 
merged to the core of UGC 05101 or is hidden behind the bright disk and jet. 
Majewski et al. (1993) offer a similar picture but also argue the 
object 17\arcsec\ to the southeast is a gas-poor dwarf galaxy which may have 
caused the ring and AGN.  In addition Majewski et al. (1993) 
argue that the structure of UGC 05101 could be the result of two 
interactions, the first causing the AGN and tidal tail and the second causing 
the ring. Carico et al. (1990a). These authors suggest a companion 50\arcsec\ 
to the west but do not mention the object claimed as a galaxy by Majewski et al. 
(1993).  Radio maps (\cite{CHSS}; \cite{Con88}) show a weaker 
component object $\sim$45\arcsec\ to the northeast of the main core of UGC 
05101, which has no counterpart on the DSS frame (Figure~\ref{fU5101}) or on the 
deep images of Sanders et al. (1988) .

To within our errors we did not resolve the inner core of  UGC 05101 (detectors 
4,5,6) in the FIR, consistent with the small ($<$ 5\arcsec) sizes seen in the 
radio (\cite{CHSS}, \cite{Sop91}).  However, we note in Figure~\ref{fU5101} an 
apparently significant excess of flux in detector 7, on the west side of the 
galaxy, as well as a possible slight deficit of flux at the center compared to 
the IRAS flux. These detectors lay along the optical jet or tidal tail that 
extends to the west, and detector 7 is where that tail crosses the ring. The 
signal in detector 7 alone could correspond to a source with a total flux of 
$5.8 \pm 1.3$ Jy.  At the distance of UGC 05101 ($\sim$160 Mpc, \cite{CHSS}) 
this corresponds to a 100 \micron\ luminosity ($\nu L_{\nu}$) of $\sim 10^{11} 
L_{\sun}$. The optical images of Sanders et al. (1988) show no optical concentration at this position. The K-band images of Genzel (1998) do not include this position in their field of view.

Since we did not resolve UGC 05101 we used the minimum size $D_{e} < 12$\arcsec, 
the IRAS fluxes, and several FIR and sub-millimeter points (\cite{Rig96a}; 
\cite{Car92}) to produce a $T_{d} \sim 34.7$, a $\tau_{100} > 1.3 \times 10^{-
2}$, an $A_{V} > 8.5$, and n $\sim$ 1.50. The galaxy q-value of 2.10 is slightly 
low as might be expected for a galaxy with a radio weak AGN undergoing a large 
starburst.  The tail seen in the optical and in the FIR does not seem to have a 
radio counterpart.  So the evidence for an extra FIR source associated with this 
tail makes the low q-value for this galaxy even more difficult to explain.

\subsubsection{NGC 3110} \label{N3110} 

NGC 3110 is a low inclination spiral with a companion about 2\arcmin\ to the 
southwest near the end of one of its two distinct spiral arms 
(Figure~\ref{fN3110}).  The system has a log $(L_{FIR}/L_{\sun}) = 10.96$ and a 
rather large gas mass (as estimated from CO detection by Sanders et al. (1991) of $2 \times 10^{10} M_{\sun}$.  This galaxy presents a fairly unique opportunity.  Since it is not too far away (65 Mpc, \cite{CHSS}), it subtends a substantial angle across the sky ($>$ 60\arcsec\ $\times$ 30\arcsec) but it is still very luminous. It provides a good opportunity to resolve a fairly typical FIR luminous galaxy.

We clearly resolve the core of NGC 3110 (Figure~\ref{fN3110}) and derive a 
$D_{g} \sim 24 $\arcsec\ and a $D_{e} \sim 13$\arcsec.  However, our photometry 
strongly suggests even further extension.  Our detector 5 flux (corrected for 
minor miscentering) and sum of detectors 1-10 flux both are $\sim$33\% below the 
IRAS flux (Table 4). This suggests that not only the core that we resolved is 
extended, but the FIR emission area is extended enough that a substantial 
portion of the FIR flux escaped of our detector array entirely.  From the flux 
received in detectors 11-20 (not shown here, perhaps 5 Jy) 
and assuming a similar amount of FIR emitting area is spread on the other side 
of the detector 1-10 arm (to the east), much of the flux discrepancy can be 
accounted for. The almost exact coincidence of the IRAS source with NGC 3110 
suggests that little of the IRAS flux comes from the companion to the southwest.  
So we conclude that the core of NGC 3110 is extended at least on a scale of 
13\arcsec, and beyond the core the disk or arms of the galaxy contribute up to 
35\% of the total FIR flux.

Using the derived sizes of the core, our array-summed 100 \micron\ flux, and a 
60 \micron\ flux scaled so that the $S_{100}/S_{60}$ ratio remains the same as 
for the BGS fluxes as well as a 1.25 mm point (\cite{Car92}) scaled 
similarly, we obtain a core dust temperature of 32.1 K and a rather high $A_{V}$ 
of 8.3. The \cite{CHSS} radio flux includes both the core and the disk so we 
only have the system q of 2.22 using BGS fluxes.

\subsubsection{NGC 4151} \label{N4151} 

Despite the fact that NGC 4151 has a Seyfert 1 active core, it is more 
representative of ``normal" nearby galaxies because of its relatively low 
luminosity of log$(L_{FIR}/L_{\sun}) = 9.5$ (using D = 17 Mpc, \cite{Hun92}).  
The source has an optical size of about 4\arcmin\ $\times$ 3\arcmin\ on the red 
Palomar Sky Survey plate (\cite{Nil73}). An ionization cone, marked by a narrow 
emission line region, occupies the central ten arcseconds, and is extended along 
an axis roughly perpendicualr to the stellar disk. We included NGC 4151 in our 
target list to confirm earlier observations of large extension on the scale of  
$\sim$100\arcsec\ at 155 \micron\ (\cite{Eng88}) and 100 \micron\ 
(\cite{Gaf92}).

Somewhat surprisingly, the extension is not obvious in the FIR profiles 
(Figure~\ref{fN4151}), which offers a stark contrast against the profiles of the 
high luminosity galaxies. We do not resolve the core but, even more so than for 
NGC 3110, we do not recover the entire IRAS flux either. This suggests that much 
of the flux is extended beyond the effective beam of our detectors.  Our 1-10 
detector flux sum is $\sim$4.1 Jy which leaves $\sim$4.5 Jy of the large-beam IRAS flux unaccounted for.  If this flux were distributed evenly over the optical disk ``ellipse" seen in Figure~\ref{fN4151}a of $\sim$150\arcsec\ $\times$ 75\arcsec\ we would be insensitive to the remaining flux because of the baseline we removed from our profiles.

Added confidence in this picture of a compact far infrared core containing about 
half the total flux surrounded by a very extended envelope comes from ISOPHOT 
observations of this object (\cite{Rod96}) which give a C100 ISOPHOT flux of 5-6 Jy at 100 \micron\ in 43 \arcsec\ pixels, compared with the 8.6 Jy IRAS flux. 

Thus we do not directly detect disk emission but only infer it from the 
differences in flux.  If the central point source were not present, we would not 
have detected the galaxy at all!  Our observations are not sufficient to 
determine if the excess IRAS flux is from the inner 100\arcsec\ as suggested by 
the evidence for extension in the Engargiola et al.(1988) data,  but they do 
not conflict with that suggestion.

The unresolved core flux of 4.1 Jy might be due to direct (non-thermal) emission 
from the AGN, but an explanation based on distributed sources from the disk with a diameter of $<$ 1 kpc is more likely in view of the fact that the spectral slope into the submillimeter is steeper than the $\alpha \sim 2.5$ which is considered the steepest slope possible from sychrotron-self absorption (\cite{Eng88}; \cite{Ede88}).  Whatever the source of flux in the unresolved core, about half of the total luminosity of NGC 4151 originates outside of that unresolved core.  Little of this extended FIR flux can be due to dust heated by the AGN.  

Assuming the bolometric energy production of the galaxy is about twice that 
which eventually comes out in the FIR from AGN heated dust, our model (a copy of 
that of \cite{Bar87}) suggests that dust heated by the AGN should produce 
emission on scales of $\sim$28\arcsec\ or greater if the dust geometry were 
such that there was dust at this distance with a unobscured view of the AGN in 
the center.  We would have detected emission on this scale if dust 
heating by the AGN was a significant source of the FIR emisison. ISO 
observations (\cite{Rod96}) appear to confirm that the FIR flux is thermal 
emission from dust heated by stars. The FIR data fit a blackbody of T = 36 K 
while a warmer dust component heated by the AGN has T = 170 K. 

It is unlikely that the AGN dominates the the FIR emission either from its 
direct emission or from the dust that it may heat, but the AGN does make its presence clear in the radio. The comparatively strong radio flux from the AGN 
(\cite{Con87}) skews the q-value to a very low value of 1.4 when we use the 
IRAS fluxes. 

\subsubsection{UGC 08696} \label{U8696} 

UGC 08696 (also known as Markarian 273) is a well observed galaxy with an 
optically conspicuous jet or tail extending to the south for at least 1\arcmin\ 
(Figure~\ref{fU8696}). Sanders et al. (1988) included it in their 
sample of ultra-luminous IR galaxies because of its high FIR luminosity of 
log$(L_{FIR}/L_{\sun}) = 11.9$. It is one of the most luminous galaxies in our 
sample. Veilleux et al. (1995) classify the nuclear source as a 
LINER from the optical emission line ratios but it is called a Seyfert 2 
elsewhere (e.g. \cite{Kha74}). 

UGC 08696 contains an OH megamaser (\cite{Sch87}) which corresponds spatially to 
the main optical component and main radio component (\cite{CHSS}, \cite{Sop91}).  
There are two radio components aligned NW-SE and separated by $\sim$1\arcsec\ 
with the NW component dominating.  In the NIR however, the SE component 
disappears and is replaced by a component to the SW (\cite{Maj93}; Zhou et al. 
1993; \cite{Car90b}).  In addition, Sanders et al. (1988) mention that there is a companion galaxy 40\arcsec\ to the north of the main component in the direction opposite the jet. This companion appears starlike on the DSS 
frame.

We do not significantly resolve the core of UGC 08696 directly at 100 \micron\ 
(Figure~\ref{fU8696}) but we do not quite recover the full IRAS flux of 22 Jy 
(BGS) in a point source either. The sum of our detectors 1-10 gives a 
flux of $17 \pm 3$ and the detector 5 flux is a similar $18 \pm 2$. The optical 
tail also emits in the radio (\cite{CHSS}) and so the additional far infrared 
flux seen in the IRAS data may be located there. Our second bank of detectors 
crosses nearly over the companion galaxy, but we do not see a recognizable local 
maximum in the far infrared emission at that position. While the excess emission 
can plausibly originate in the jet, it is noteworthy that Turner, Urry and 
Mushotsky (1993) show a bright serendipitious x-ray source that is 
approximately coincident with the starlike object at the DSS image in Figure 10. 
Neither of these regions are sampled in our data. On the other hand, a component 
of emission that is spatially offset from the main peak might be expected to 
displace the IRAS peak. The IRAS peak is, however, coincident with the radio 
source.

Using our 100 \micron\ flux and a 60 \micron\ flux scaled so the ratio of 
S(60)/S(100) is the same as in the BGS we obtain, with the \cite{CHSS} 1.49 
GHz radio flux, a q-value of 2.16 which is comparatively low.  Using the IRAS 
fluxes the q-value becomes 2.27, much closer to the expected $2.34 \pm 0.2$ 
(\cite{Con91a}).  In addition, using other sub-millimeter to millimeter fluxes 
(\cite{Rig96a}; \cite{Kru88}) the best fit single temperature dust model gives a 
dust emissivity exponent n = 1.35 when we use our observed 100 \micron\ flux and 
the scaled 60 \micron\ point.  When we use the \cite{BGS} fluxes, the exponent n 
= 1.51.  In general, astronomical dust is thought to have an n = 1 - 2 
(\cite{Car92}), but for every other galaxy for which we can calculate an 
exponent n, we generally obtain a value near or greater than 1.5 (Table 8).

The low q-value and low n we obtain when we used our fluxes suggests the higher 
IRAS 100 \micron\ flux is a better match to the other wide beam (full system) 
measurements. Simple subtraction of the our detector 1-10 flux sum from the IRAS 
flux gives a possible flux for the tail of $5.1 \pm 2.9$ which corresponds to a 
luminosity ($\nu L_{\nu}$) of $1.2 \times 10^{11} L_{\sun}$. 

Using the BGS fluxes and the upper limit of 12\arcsec\ for the FIR 
diameter of the emitting region (along with the sub-millimeter fluxes mentioned 
above) we obtain a best fit dust temperature of 42.0 K and an optical depth at 
100 \micron\ of $> 6.4 \times 10^{-3}$ which implies an $A_{V} > 4.8$.

\subsubsection{NGC 6090} \label{N6090}

NGC 6090 is a pair of galaxies separated by $\sim$10\arcsec\ and aligned NE-SW (Figure~\ref{fN6090}a).  According to  Martin et al. (1991) the optical cores of the two galaxies are in contact and the system also includes ``wings" (barely visible in the DSS frame) which makes the pair reminiscent of the famous ``Antennae" system.  The entire system has an optical size 2.8\arcmin\ $\times$ 1.5\arcmin\ (\cite{Nil73}) although our DSS image (Figure~\ref{fN6090}) shows that the optical cores have a combined size of 20\arcsec\ $\times$ 10\arcsec\ with a larger size around the northern dominant galaxy. 

Despite the fact that our array was aligned along the NE-SW line between the two galaxies, we did not resolve NGC 6090 (Figure~\ref{fN6090}). Essentially the entire IRAS 100 \micron\ flux is recovered in our beam. The approximate point source we observed is skewed towards detector 4 as we might expect if the southwestern source were responsible for a significant part of the emission, but due to our pointing errors, this cannot be confirmed.  The radio map in \cite{CHSS} shows that the northeastern source is dominant in the radio.  Our small size ($D_{e} < 12$\arcsec) is consistent with the radio size of the system given by \cite{CHSS} as 5\arcsec\ $\times$ 7\arcsec\ .
 
We note that Hwang et al. (1999) used ISOCAM images to detect some 
small extension of NGC 6090 at 15 \micron\, but their peak-to-total flux 
analysis did not provide quantitative spatial information that can be compared 
to our data. It is also noteworthy that Bushouse, Telesco, and Werner (1998) found little or no evidence for extension of this source at 10 \micron\ from ground-based data. 

ISO photometry (\cite{Aco96}) has given the energy distribution of NGC 6090 from 
3.6 to 200 \micron\ and shown that the galaxies are in fact dominated by 
starburst heated dust. The authors of this work suggest without elaboration that 
the source is resolved with ISOPHOT in the 60 \micron\ band, though other 
galaxies in the field may have contributed to this impression. Acosta-Pulido et 
al. (1996) also suggest a second, subsidiary dust component with 
$T_{d} = 20$ K, but they do not use the 1300 \micron\ measurement of Chini et 
al. (1992) because of the small beamsize (11\arcsec) used for this 
measurement compared with the ISO beamsize.

Since the angular size we derive is $< 12$\arcsec\, we feel confident in using 
the 1300 \micron\ point with out data.  Our model (without the ISO data) then 
gives $T_{d} = 31 K$, $\tau_{100} > 8.9 \times 10^{-2}$, and n $\sim$ 2.3. 

\subsubsection{NGC 6286} \label{N6286} 

NGC 6286 is an edge-on spiral in an interacting pair with the spiral NGC 6285 
$\sim$1.5\arcmin\ to the northwest (Figure~\ref{fN6286}).  The optical 
extension of NGC 6286 is 1.3\arcmin\ $\times$ 1.2\arcmin\ (\cite{Nil73}).  
Unfortunately, the long axis of our array was aligned almost perpendicular to 
the long axis of the galaxy and we do not get nearly as much information about 
FIR emission along the disk as we might have had if we had a more favorable 
array alignment.  Filaments and plumes seen in deep CCD images as well as a 
faint shell-like feature extending about 0.5\arcmin\ to the ESE of NGC 
6286 (barely visible in our DSS image) cause Whitmore et al. (1990) to brand NGC 6286 a possible polar ring galaxy. 

Despite the poor position angle, we may have resolved NGC 6286 along the polar 
axis (Figure~\ref{fN6286}).  The flux in detector 3 is well above the PSP, and 
we do not recover the IRAS flux in a point source. The  $D_{g} = 21$\arcsec\ 
while the $D_{e} = 12$\arcsec. The sum of detector 1-10 
fluxes match the IRAS flux well.  The fact that we still resolve the emission 
although the galaxy disk is not aligned with our array suggests that the extended FIR emission is likely to be associated with some non-disk component. In this respect, we note that detector 3, which sits on the shell-like feature to the ESE is well above the point source profile. Using the size of 12\arcsec, our flux for detectors 1-10 at 100 \micron\, and a flux from Surace et al. (1993) for NGC 6286 which excludes the contribution from NGC 6285 at 60 \micron\, we get the lowest dust temperature of our sample of $T_{d} = 28.6$ (when we also use the 1.25 mm value from Carico et al. (1992) to get n = 1.77).  Our optical depth measurement of $\tau_{100} = 3.2 \times 10^{-2}$ translates into an $A_{V} = 24$. 

Aside from the large deviation from a point source in detector 3, the FIR emission can be attributed to a nuclear bulge, leaving little evidence for disk emission. Radio maps (\cite{CHSS}; \cite{CHYT}) show only slight emission from the disk as well. 

The q-value for NGC 6286 is 2.01 which is slightly lower than those of most 
other galaxies.  Our higher resolution observations confirm that NGC 6285 is not responsible for the anomaly. Lonsdale et al. (1993) classify NGC 6286 (misnamed NGC 6285 in their paper) as \ion{H}{2} spectral type, but Veilleux et al. (1995)  classify it as having a LINER spectrum so it is possible that NGC 6286 contains a weak AGN core which adds extra radio flux and lowers the q-value from the strong starburst.

\subsubsection{IRAS 17132+5313} \label{I1713} 

IRAS 17132+5313 is a double galaxy aligned E-W with the eastern component 
brighter in the visible and radio (Figure~\ref{fI1713}).  The two components 
are separated by about 10\arcsec.  Interestingly, most studies of the system in 
the radio have concentrated on the weaker western component because it is very 
compact.  The eastern component was resolved by \cite{CHSS} with a size of 
3\arcsec\ $\times$ 1.6\arcsec\ and it has an \ion{H}{2} like spectrum 
(\cite{Vei95}).  The western component is resolved by both \cite{CHYT} and by  
Lonsdale et al. (1993) who classify the western source as an AGN.

We did not resolve the galaxy at 100 \micron\ (Figure~\ref{fI1713}).  The IRAS flux is slightly higher than our value but is within the errors.  The 
companion galaxy about 1\arcmin\ to the southeast of the dominant pair may 
contribute some significant flux to the IRAS value since it emits $\sim$13\% of 
the total radio flux. 

We model IRAS 17132+5313 with the full IRAS fluxes from the BGS and use 
the unresolved upper limit of 12\arcsec.  The model provides a dust temperature 
of 37 K and an optical depth at 100 \micron\ of $>$ 3.6 $\times 10^{-3}$ which 
corresponds to an A$_{V}$ of $>$ 2.7.  The FIR flux along with the total 
\cite{CHSS} 1.49 GHz radio flux of 25.8 mJy give a q-value of 2.51.

\subsubsection{IRAS 17208-0014} \label{I1720} 

IRAS 17208-0014 is an extremely luminous and distant system which like VII Zw 31 
seems to have a large amount of gas concentrated in a very small volume. It is 
formally the most luminous source in our sample, and provides one of our more 
surprising results. From single dish observations of the CO, Mirabel et al. 
(1990) obtain a molecular gas mass of $5.5 \times 10^{10} M_{\sun}$. Observations with the Owens Valley interferometer array geta similar amount.  The CO emission is spatially unresolved and so has a scale size of $<$ 3\arcsec.  Observations in the CO J(2-1) line (\cite{Rig96b}) provide a smaller estimate for the gas mass of only $3.8 \times 10^{9} M_{\sun}$.

IRAS 17208-0014 also contains a OH megamaser (\cite{Mar89}) suggesting a cloud 
in front of a continuum source, and so it is not surprising that \ion{H}{1} is strong in absorption. The \ion{H}{1} column density is $1.7 \times 10^{22}$ cm$^{-2}$ if a spin temperature of 100 K is assumed.

Our 100 \micron\ fluxes for IRAS 17208-0014 are $19.4 \pm 4$ Jy (sum of 
detectors 1-10) and $24.8 \pm 3$ Jy (detector 5).  These are both far below 
the BGS2 flux of 38 Jy.  The spatial distribution shows no 
particularly great deviation from a point-source at 100 \micron\ 
(Figure~\ref{fI1720}).  The DSS image shows a fairly crowded field, but no 
extended companions which could be responsible for the extra flux are apparent 
(Figure~\ref{fI1720}).  There are no apparent companions in the K-band either 
(\cite{Zen93}; \cite{Mur96}). We could find no radio observations which might 
point out any existing companion in the field.  We do note that in the deep red band image of Murphy et al. (1996) an arm extends to the east about 30\arcsec\ from the core; it may be that IRAS detected FIR emission from this arm and our smaller beam did not. This image also shows what appears to be a bridge of emission connecting the galaxy with the starlike object 0.5 \arcmin\ to the north, and it is possible that this bridge and separate source, which are not well sampled by our array, accounts for some of the missing energy. 
But these explanations are complicated by the fact that that the IRAS centroid 
is identical to ours, suggesting that any missing flux should be symmetrically 
distributed around the center. Another possibility is that the FIR flux of this 
source is variable but  radio observations at 4.85 GHz show no significant 
variation from 1987 to 1992 (\cite{Bec91}; \cite{Gri95}; Condon, Anderson, \& 
Broderick 1995), and the source spectrum is clearly thermal throughout the infrared.

With the 1.425 GHz radio flux of 102 mJy from Condon et al. (1995) 
and the 4.86 GHz flux of 61 mJy from Condon et al. (1996) we obtain a power law slope of $\alpha = -0.42$  ($S_{\nu} \propto \nu ^{\alpha}$) and a 1.49 GHz value of 100 mJy from which we obtain a q-value of 2.62 when using the BGS2 FIR fluxes.  The q-value with the BGS2 FIR fluxes and the 4.86 GHz value is 2.85.  Both of these q-values are somewhat high. Summing the flux across our array, and using  a 60 \micron\ value such that the ratio between the 60 and 100 \micron\ points matched the ratio of the BGS2 values, we obtained much more normal q-values of 2.33 and 2.56 for 1.49 GHz and 4.85 GHz respectively. See Section~\ref{qval} for a brief discussion on the possibility that OH megamaser galaxies have high q-values.

With its very compact size (unresolved with 3\arcsec\ beam in CO), the simple 
modeling of IRAS 17208-0014 (with our 100 \micron\ flux, and the IRAS 60 
\micron\ flux scaled so as to match the \cite{BGS2} flux ratio) gives a very 
high optical depth of $\tau_{100} > 9.3 \times 10^{-2}$ which corresponds to an 
$A_{V} >$ 70.  This is by far the highest optical depth we have calculated for 
our sample.

\subsubsection{UGC 10923} \label{U10923} 

UGC 10923 (Mkn 1116) is a multiple system with an interacting companion only 
15\arcsec\ to the northeast of the main component. The DSS frame shows an 
apparent bridge connecting these components to a compact object 20\arcsec\ to the northwest. Another galaxy lies 50\arcsec\ to the southeast. Our data shows 
conclusively that most of the FIR flux comes from the cluster galaxies on the 
west side(Figure~\ref{fU10923}a).  Because of its low 60 \micron\ flux UGC 10923 is not in the BGS and so is less often observed than other FIR luminous 
galaxies. The object is poorly documented in the literature. Bushouse (1987) 
observed UGC 10923 at the 21 cm line in \ion{H}{1} and estimates a atomic gas mass of $7 \times 10^{9} M_{\sun}$.  In addition, Bushouse provides an H$\alpha$ image which shows a size of $\sim$20\arcsec\ in the western group.
 
From the comparison to the point source function (Figure~\ref{fU10923}) it is 
clear that we resolved UGC 10923 at 100 \micron\ . The H$\alpha$ size is well matched by the 20\arcsec\ size of our resolved Gaussian model ($D_{g}$). Summing over all our detectors, we recover the IRAS flux 
(\cite{Maz91}) to within our errors. 

Away from the central source, the emission appears to be distributed mostly 
eastward the more distant component. That component itself does not appear to be a significant contributor to the 100 \micron\ flux. The IRAS centroid is offset from the main component slightly toward the east, which is consistent with our 
findings. 

Using the IRAS fluxes, an additional upper limit of 5.7 mJy at 1300 \micron\ 
from Chini, Kr\"{u}gel, \& Kreysa (1992), and the $D_{e} = 11$\arcsec\ size of the region, we calculate a dust 
temperature $< 28$ K, an optical depth at 100 \micron\ $> 2.1 \times 10^{-2}$, 
and a dust emissivity exponent n $> 2.2$.  This value for n is nearly as high as 
any other emissivity exponent for dust we calculate and it is higher than the 
normally accepted estimates of n = 1 - 2 for dust (\cite{Car92}) but its low 
value depends entirely on the low upper limit given by Chini et al. (1992). The $\tau_{100}$ corresponds to an $A_{V} > 16$ so the H$\alpha$ is prbably leaking out of the starburst region. 

We found no published radio fluxes at or near 1.49 GHz for UGC 10923 but
Marx et al. (1994) 
 give fluxes at 4.76 GHz and 10.7 GHz from which we derive a flux at 4.85 GHz.  
When combined with the IRAS fluxes the derived radio flux provides a normal q-
value (for 4.85 GHz) of 2.68.

\subsubsection{NGC 7469} \label{N7469} 

NGC 7469 is a bright, luminous, and well observed SBa galaxy with a Seyfert 1 
core (\cite{Gen95}, \cite{Cut84}). The companion IC 5283 is about $\sim$80\arcsec\ to the north.  NGC 7469 has an optical size of 100\arcsec\ $\times$ 60\arcsec\ (\cite{Nil73}) but our DSS frame only shows the core with a size of $\sim$40\arcsec\ $\times$ 20\arcsec\ (Figure~\ref{fN7469}).  Near IR 
profiles show a sharply peaked distribution with scale size of 
$\sim$2\arcsec\ (\cite{Zen93}) or less (\cite{Ter94}; \cite{Maz94}; 
\cite{Gen95}).  There is $1.5 \times 10^{10} M_{\sun}$ of $H_{2}$ concentrated 
in the central 2\arcsec\ of NGC 7469 as determined by studies of CO emission 
(\cite{Mei90}) and somewhat less of \ion{H}{1} (\cite{Mir88}).  Some of this gas 
is concentrated in a clumpy nuclear ring of radius $\sim$1.5\arcsec\ seen in 
radio (\cite{Wil91}; \cite{CHYT}), visible light (\cite{Mau94}), and even at 
11.7 \micron\ (Miles, Houck, \& Hayward 1994). In addition, NGC 7469 shows emission lines from polycyclic aromatic hydrocarbons (PAHs) from the extra-nuclear region (\cite{Mil94}; \cite{Maz94}; \cite{Cut84}). Miles et al. (1994) point out that these molecules would be destroyed by the X-ray flux from the active core of NGC 7469 and note that shielding by clumps of gas with N$_{H} > 10^{23}$ atoms cm$^{-2}$ could protect them.

Despite the assertion that the majority of the log$(L_{FIR}/L_{\sun}) = 11.2$ 
is concentrated within the circumnuclear ring with a $\sim$2\arcsec\ diameter 
(\cite{Gen95}) we may have see a very small fraction of 100 \micron\ flux on a considerably larger scale (Figure~\ref{fN7469}). Our fitting and deconvolution on this high S/N object allows some superresolution, and gives $D_{g}=14$\arcsec\ and $D_{e}$ = 7\arcsec. While the majority of the FIR flux lies on scales well below our resolution there may be some flux extended beyond the circumnuclear ring.  Photometrically, we recover exactly the IRAS flux, within our errors and so cannot use a flux deficit to estimate the size.

With our estimate of $D_{e} = 7$\arcsec, the single slab model for the dust 
gives $\tau_{100} > 5 \times 10^{-2}$, which we convert to an $A_{V} > 35$. This extinction cannot surround the AGN since it is easily visible in 
the optical.  We note also that using N$_{H}/A_{V} = 1.9 \times 10^{25}$ 
(\cite{Boh78}) we get the column density N$_{H} > 6.7 \times 10^{22}$, which is 
high enough to adequately shield the PAHs (\cite{Mil94}).

The flux we obtained for NGC 7469 used only detectors 1-9 for both the baseline 
and the flux by summation of detectors.  We did this because of the excess flux 
seen by detector 10, which is close to IC 5283. While the IRAS position does not 
appear to be biased by this companion, and the IRAS flux was recovered in 
detectors that excluded it, our data is suggestive that IC 5283 contributes 
modestly to the luminosity of the system. These two galaxies were resolved 
separately at 12 and 25 \micron\ with HIRES studies on the IRAS database 
(\cite{Sur93}). In this study, IC 5283 was found to be less than 5\%\ of NGC 7469 at 25 \micron\ -- a ratio that is surprisingly small given the conspicuousness of the former in our dataset, in which it is probably only partly sampled. It would appear that IC 5283 has a cooler spectrum than NGC 7469. 

Optical studies of this pair (\cite{Marq94}) show that IC 5283 is, in itself, a 
strongly disturbed system. There appears to be little evidence for a substantial 
stellar or gaseous component bridging the systems, and the lack of 100 \micron\ 
emission between them (in our detectors 7-9) is therefore not surprising.

\subsubsection{NGC 7541} \label{N7541} 

NGC 7541 (Figure~\ref{fN7541}) is a relatively nearby (36 Mpc, \cite{CHSS})  
disturbed starburst spiral with an optical size of 3.4\arcmin\ $\times$ 
1.1\arcmin\ (\cite{Nil73}).  Radio maps by \cite{CHSS} and  Colbert et al. (1996) also give a rather extended size of 60\arcsec $\times$ 24\arcsec, with little central concentration. At 10 \micron\ the small aperture (5\arcsec) to large aperture (100\arcsec) compactness ratio is C = 0.07 (\cite{Giu94}). The non-point-like flux distributions at radio and 10 \micron\ wavelengths suggest that the FIR flux of NGC 7541 will also be extended.

NGC 7541 has been observed often in \ion{H}{1} (\cite{Lu93}; \cite{Oos93}).  
We use a value of 47 mJy which implies an atomic gas mass of $9 \times 10^{9} 
M_{\sun}$ (\cite{Mir88}).  This is approximately twice the molecular gas mass of 
$4.5 \times 10^{9} M_{\sun}$ (\cite{San86}).  The usual ratio of molecular to 
atomic gas masses in FIR luminous galaxies is $\sim$2-5 (\cite{Mir88}) and so 
in this respect the ratio of $\sim$1/2 makes NGC 7541 look more like a quiescent galaxy.

We easily resolve NGC 7541 at 100 \micron\ along the minor axis of the galaxy 
(Figure~\ref{fN7541}).  Our fitted Gaussian gives $D_{g} = 23$\arcsec\ while 
the fitted disk model gives a $D_{e} = 12.5$\arcsec.  A note in \cite{CHSS} 
says the companion NGC 7537, which lies several arcminutes to the southwest, may 
contribute to the IRAS flux.  Our summed flux ($38.7 \pm 6.0$ Jy) is consistent 
with the IRAS flux of 40.6 Jy, so we have recovered the entire IRAS flux despite 
the alignment of our array along the minor axis of NGC 7541. These facts taken 
together would argue that the morphology of the 100 \micron\ emission is more 
circular than the very elongated optical contours. If distributed like the 
optical emission, we would have missed most of the flux in our array.

We note parenthetically that our array position did not cover the site of the 
recent supernova 1998d, a SNIa which was discovered 50"W and 10"N of the center. 

Using the single slab model, IRAS fluxes, and the 1.25 mm flux from Carico et 
al. (1992) we get $\tau_{100} = 3.4 \times 10^{-2}$. This gives $A_{V} = 26$ which is not unreasonable for an edge-on spiral.

\subsubsection{Zw 475.056} \label{Z475} 

Zw 475.056 (IC 5298) (Figure~\ref{fZ475}) is a Seyfert 2 galaxy (\cite{Vei95}) 
which has an associated OH maser (\cite{Mir87}).  The molecular gas shows a 
double peaked line with a strength that corresponds to a mass of $9.4 \times 
10^{9} M_{\sun}$ (\cite{Mir87}). The double peaked line is also apparent in the 
\ion{H}{1} observations and is typical of galaxies which have not undergone 
interactions.  The \ion{H}{1} line strength suggests an atomic gas mass of $5.5 
\times 10^{9} M_{\sun}$ (\cite{Mir87}).  Considering the fact that all the other 
OH megamaser sources in the sample show at least some \ion{H}{1} absorption, the 
double peak structure of the gas lines here may be due to foreground absorption 
of the galaxy (\cite{Mir87}). If it is an absorption feature, the gas masses 
quoted above are only lower limits.

Our profile of Zw 475.056 (Figure~\ref{fZ475}) does not show significant 
evidence for extension and our 100 \micron\ point source fluxes matches the IRAS 
BGS flux well.  Using $D_{e} < 12$\arcsec\ we derive a dust temperature 
of 38 K and an optical depth at 100 \micron\ greater than $5.0 \times 10^{-3}$ 
which corresponds to an $A_{V} > 3.8$. That the IRAS position is significantly 
displaced from the place where we see all the flux, and no companions that might 
confuse the centroiding are evident, has no obvious explanation. 

VLBI observations of Zw 475.056 (\cite{Lon93}) resolve the core as do the VLA observations in \cite{CHYT}. These latter data show that the core of the galaxy, on scales less than an arcsecond, is oriented NNW-SSE as in the larger scale optical (Figure~\ref{fZ475}) and I-band (\cite{Zen93}) contours.  The faint halo with $\sim$20\arcsec\ extent visible in the optical (see the DSS image Figure~\ref{fZ475} is not visible in the I, H and K. These images show only a core less than 10\arcsec\ in diameter (\cite{Zen93}). This halo evidently does not contribute much of the 100 \micron\ emission. The radio flux for Zw 475.056 at 1.49 GHz (\cite{CHSS}) provides a q-value of 2.53 which is somewhat above the average value of 2.34.  A flux at 4.85 GHz (\cite{Sop92}) gives a q-value of 2.97 which is also higher than normal.  See the discussion on OH megamasers having high q-values in Section~\ref{qval}. 

\subsubsection{NGC 7625} \label{N7625} 

NGC 7625 is a type Sa/S pec galaxy with a comparatively modest luminosity. 
Unlike most early type galaxies, it happens to have a great deal of gas and dust 
associated with it and large star formation rate (\cite{Li93}). A dust lane is 
conspicuous in the DSS image (Figure~\ref{fN7625}).  The optical size of the 
galaxy extends to 1.5\arcmin\ $\times$ 1.5\arcmin\ (\cite{Nil73}) but 
comparisons of the sizes of the blue light, the CO, the H$\alpha$, and the 20 cm 
emitting regions along the major axis of rotation (PA $\sim$ 28\arcdeg) show 
that each of these tracers give a FWHM of $\sim$10\arcsec\ (\cite{Li93}). 
NGC 7625 has a molecular gas mass of $\sim 2.4 \times 10^{9} M_{\sun}$ and an 
atomic mass of about the same amount (\cite{Li93}).

Our 100 \micron\ profile (Figure~\ref{fN7625}) is peaked between detectors 4 and 5, while the optical peak should have been between detectors 5 and 6. This suggests that the FIR is skewed slightly to the south away from the optical (\cite{Li93}) and large beam 1.49 GHz radio (\cite{CHSS}) position which we attempted to center up on. That pointing position also corresponds well to the H$\alpha$ and \ion{H}{1} center while the CO and small beam radio centers appear to be several arcseconds to the northwest. Significantly, the FIR peak corresponds spatially better with the prominent dust lane, and the centroid of the outer optical contours. 

Though our point source flux is consistent with that of IRAS, there is some 
slight evidence for extension of NGC 7625 in the profile, (Figure~\ref{fN7625}) 
and we obtain a $D_{g} = 21.8$\arcsec\ and $D_{e} = 13$\arcsec\ which matches 
the sizes of the other tracers of the starburst activity.  Using the derived 
disk size of 13\arcsec\ in the single slab model, we derive a $T_{d} = 32$ K 
and a $\tau_{100}= 1.2 \times 10^{-2}$ which translates into an $A_{V} = 9$.

\subsubsection{NGC 7770/7771} \label{N7771} 

NGC 7770/7771 is an interacting system with the larger spiral galaxy NGC 7771 
separated from its companion to the southwest by about 80\arcsec\ 
(Figures~\ref{fN7770}; ~\ref{fN7771}). NGC7771 is a luminous system 
that contains a well defined starburst ring (Smith et al. 1999) with a major axis of 6\arcsec\. The optical morphology is strongly affected by extinction. We found that both galaxies are emitting in the FIR although the system is dominated by NGC 7771. Both galaxies also have optical spectra classified as \ion{H}{2} emission-dominant (\cite{Kim95}; \cite{Vei95}).

We did not spatially resolve the core of NGC 7771 (Figure~\ref{fN7771}), which 
argues that essentially all of the far infrared emission comes from within the 
starburst ring, the diameter of which is just below our detection limit for 
spatial extension. Our observations of NGC 7770 are too noisy to determine a 
size (Figure~\ref{fN7770}), so we do not produce a model to determine its 
optical depth. Our observations of NGC 7770 do suggest that it may not be a 
point source, however, and the I-band image shown in \cite{Smi99} shows a 
20 \arcsec extent that has little central condensation.

Using BGS fluxes (with both the 100 and 60 \micron\ fluxes scaled by the 
radio ratio) for NGC 7771 as well as the 1.25 mm point from Carico et al. (1992) 
we obtain a dust temperature of 33 K. The model also gives an optical depth at 100 \micron\ $\tau_{100} >$ 2.3 $\times 10^{-3}$ and a dust emissivity exponent n $\sim$ 1.49.  The $\tau_{100}$ corresponds to an $A_{V} > 17$. 

Using the radio fluxes from \cite{CHSS} and BGS fluxes give a normal q-value of 2.39 for the system.  We note here that despite its small flux compared 
to its companion, NGC 7770 is fairly luminous by itself with 
log$(L_{FIR}/L_{\sun})= 10.20$.  This is enough to classify it as a FIR luminous 
galaxy and help to confirm the suggestion that in interacting galaxies, both 
partners are often FIR enhanced (\cite{Sur93}, \cite{Ber93}).

\subsubsection{Markarian 331} \label{Mk331} 

Markarian 331 (Figure~\ref{fMk331}) is a FIR luminous galaxy with an HII-like optical spectrum (\cite{Vei95}).  The galaxy shows both emission and absorption in \ion{H}{1} with an atomic gas mass $> 9.55 \times 10^{9} M_{\sun}$. and column density of $7 \times 10^{20}$ cm$^{-2}$ if a spin temperature of 100 K is assumed (\cite{Mir88}).  The estimated molecular gas mass is $1.29 \times 10^{10} M_{\sun}$ (\cite{San91}).

We do not resolve Mkn 331 (Figure~\ref{fMk331}) and our flux matches the IRAS 
BGS flux to within our errors. The IRAS position is offset from the radio 
position (\cite{CHSS}) in the direction of two faint galaxies $\sim$1.5\arcmin\ 
to the southwest.  While this offset might suggest that one of the nearby companions is responsible for some of the FIR flux assigned to this object, and is introducing a bias into the large beam position centroid, our photometry does not support this explanation, and we note that the IRAS error ellipse is fairly large.

Using the \cite{BGS} FIR fluxes and a upper limit of 8.7 mJy from Chini et al. 
(1992) we estimate a dust temperature near 35 K, an optical
depth $\tau_{100} > 1.5 \times 10^{-2}$, and a dust emissivity exponent n $>$ 
2.01.  The corresponding $A_{V} > 8.6$.  The q-value with the radio value from 
\cite{CHSS} is a moderately high 2.51.

\subsubsection{UGC 12915} \label{U12915} 

UGC 12915 is the smaller of a closely interacting pair of galaxies 
(Figure~\ref{fU12915}). UGC 12914 is only 1.5\arcmin\ to the southwest.  Both 
galaxies are obviously disturbed spirals with prominent tidal tails and even a 
ring around UGC 12914.  Radio maps (\cite{Con93}; \cite{CHSS}; \cite{Con91b}) 
show not only that the galaxies are radio sources, but that the space between 
them is a radio source as well, such that only $\sim$58\% of the radio flux is 
localized around the individual galaxies.  Because it is stronger 
in the radio and perhaps at 60 \micron\ as well, we chose to center our array on UGC 12915.  The second arm of the array fell between the two galaxies allowing us to see if any of the FIR flux was coming from the radio bridge.

We resolved UGC 12915 in our first set of observations in 1994 August and found 
a $D_{g} = 24$\arcsec\ or a $D_{e} = 12$\arcsec. Our measured 100 \micron\ flux 
in detectors 1-10 was 9.9 Jy, compared to 13.4 Jy from IRAS, suggesting that 
about a quarter of the emission originates elsewhere. The second arm of the array seemed to show evidence for substantial FIR emission from the radio bridge but the uncertainty was high.

In order to confirm the photometry we duplicated the observation of UGC 12915 in 
our flight series of 1995 August.  The new observations (Figure~\ref{fU12915}, 
Table 4) confirm the size very well; they give $D_{g} = 24 $\arcsec\ and $D_{e} 
= 14$\arcsec.  The 1995 observations also match the 1994 flux estimate with a 
100 \micron\ flux of 11.6 Jy. The new observations allow for only slight FIR 
from the radio bridge.  Condon et al. (1993) claim that about 25\% of the 60 \micron\ flux is from UGC 12914. Assuming similar colors, the difference between our 100 \micron\ flux and the large beam IRAS flux from BGS (13.4 Jy) is easily explained by flux from UGC 12914.

Using the disk size of 14\arcsec, our detector 1-10 summed 100 \micron\ flux, and a 60 \micron\ flux scaled so that the S$_{60}$/S$_{100}$ ratio is the same as in the BGS, we obtain a T$_{d}$ = 31 K and a $\tau_{100}$ = 7.7 $\times 10^{-3}$ for UGC 12915.

Condon et al. (1993) had assumed that the low 1.49 GHz q-value for the UGC 12914/5 system of 1.94 occured because the radio bridge was emitting radio sychrotron but not FIR.  Our observations have confirmed that there is little if any FIR from the area between the galaxies.  Now that we have the 100 \micron\ flux for UGC 12915 we can attempt to calculate its q-value independently.  We use our 1995 August 100 \micron\ flux in detectors 1-10 and a flux for 60 \micron\ scaled so that the S$_{60}$/S$_{100}$ ratio matches the \cite{BGS} flux ratio.  With the Condon et al. (1993) 1.49 GHz flux for UGC 12915 of 47 mJy, we calculate a q = 2.26.  This normal q-value also confirms that the abnormally low system q-value is caused by the extra radio emission from the bridge and not from the disk of UGC 12915. 

\section{Discussion} \label{discuss}

Our observational techniques have been effective in probing the structure of far 
infrared emission in galaxies on very small scales. Of the 22 galaxies, we see 
at least some evidence for extension of the 100 \micron\ emission, either by 
comparison with point source profiles, or by missing peak flux, in 12 (MCG +02-
04-25, NGC 1275, UGC 05101, NGC 3110, UGC 08696, NGC 6286, IRAS 17208-0014, UGC 
10923, NGC 7541, NGC 7625, NGC 7770, and UGC 12915). The emission from NGC 4151 
appears to be very large, and is mostly outside our sampling area. 

There are several possible explanations for the strong FIR emission from the 
sample galaxies. These include:  (1) emission from an active galactic nucleus 
(AGN), (2) emission from dust heated by an AGN, (3) emission from dust heated by a massive burst of star formation, (4) emission by dust heated by an stellar 
population (5) emission by dust heated by the UV photons produced by shocks 
in galaxy collisions, and (6) emission by dust heated by extremely hot gas from a galaxy cluster cooling flow.  Clearly, not all of these mechanisms are possible in all of the galaxies in our sample.  Mechanisms (1) and (2) are only possible for those galaxies with an AGN, mechanism (5) is only possible for those systems with multiple galaxies or galaxy cores, and mechanism (6) only applies to NGC 1275 because it is the lone galaxy in our sample with a cooling flow. We will describe the expected sizes of FIR emitting regions for these mechanisms below and then compare expected sizes with our observed sizes at 100 \micron.

\subsection{Emission from AGN and Emission from Dust Heated by AGN}

If the core is resolved, little of the FIR emission can come directly from the 
point-like AGN.  For galaxies with a FIR $D_{g} > 20$\arcsec\ (barely resolved), 
we find that 50\% of the flux could come directly from a point source only if 
the remaining emission from the galaxy disk had $D_{g} > 40$\arcsec.  Radio maps 
of the galaxy disks in those galaxies we resolved (\cite{CHSS}; \cite{Con96}) do 
not show substantial emission at such scales.  To first order, radio 
distributions appear to be larger than the FIR for galaxy disks (\cite{Bic90}; 
\cite{Mar95}; \cite{Lu96}).  So a point-like AGN would provide less than 25\% of 
the FIR flux to a marginally resolved source and even less to the flux of a 
source that is more clearly resolved.

Dust heating by the AGN is another process which might produce some of the 
extended emission.  A model that assumes the dust is optically thin in the IR 
and is distributed spherically around the AGN (\cite{Bar87}) suggests that in order to match both the observed FIR luminosity and $\sim$40 K dust temperatures with a central AGN heating source, the dust must have a direct line of sight to that source (for instance if the dust was in a warped disk -- see \cite{San89}) and also must be several kiloparsecs (typically $>$ 30\arcsec) away from the AGN.  Any substantial amount of dust closer to the AGN would have a higher temperature an so would move the peak emission to shorter wavelengths.  In addition, the model AGN luminosity cannot be lowered to allow $\sim$40 K dust at smaller radii 
or else the FIR luminosity will drop below the observed levels.  Only for MCG+02-04-025 do the sizes we estimated for the FIR flux from AGN heated dust and the observed sizes match ($D_{g}\sim$19\arcsec\ for both), but it has an \ion{H}{2} like optical spectrum. Galaxies with high FIR emission tend to be complicated systems, often undergoing or having already undergone interactions AGN activity is at least as likely to be the result of the interaction as it is likely to be a determining factor in the far infrared energetics of the system. 

\subsection{Emission from Dust Heated by Starbursts}

Massive star formation is a more likely source for the energy that heats the dust in FIRLGs.  Most of the galaxies in the sample show clear evidence for widespread and intense star formation, usually a central region with an emission spectrum that resembles an \ion{H}{2} region (see Table 2). The FIR emitting region would then be expected to be similar in size to the starbursting region because the large amount of dust in typical star forming regions causes the mean free path of the  UV photons to be very small.  The expected sizes of the FIR emitting regions in this case would be variable from galaxy to galaxy but would 
correspond to the extent of photoionized gas, as traced by the emission lines. Most of our target galaxies show far infrared distributions that are consistent with the spatial tracers of massive star formation.

Starburst region sizes can be assumed to be the same as that of the H$\alpha$ regions in galaxy images if the extinction is small or sufficiently patchy.  A quick search of the literature for H$\alpha$ region sizes for galaxies we clearly resolved provides three galaxies, each with an $H\alpha$ region with FWHM $\sim$ 20\arcsec\ (NGC 7625, \cite{Li93}; UGC 10923 and UGC 12915, \cite{Bus87}).  These widths are a good match to the observed FIR sizes ($D_{g}$) listed in Table 5. So the H$\alpha$ sizes support the starburst theory of FIR emission for these resolved galaxies.

\subsection{Emission from Dust Heated by Older Stars}

It is also possible that older stellar populations provide some of the 
energy to heat the dust in luminous IR galaxies, but unless this population is hugely in excess of that indicated by the optical and near infrared tracers of the red stellar population, it is unlikely that they could dominate the energetics in the luminous systems. In particular, the NIR and FIR fluxes are not correlated in the most luminous FIR luminous galaxies (\cite{Har87}), and the colors imply insignifigant extinction towards that older population
(\cite{Har87}; \cite{Hou85}). For the lowest luminosity galaxies in our sample, the NIR luminosities are close enough to the FIR luminosities that dust heated by red giants might contribute to the FIR.  However, even these low luminosity galaxies mostly show \ion{H}{2} emission line spectra and large H$\alpha$ luminosities suggestive of dominance by a younger starburst. NGC 4151 is a noteworthy case, in that we find a large part of the far infrared emission from that galaxy to be clearly extended on a scale that is similar to the near infrared light, and over a region that has little or no ionized gas.

\subsection{Emission from Other Mechanisms}

Other mechanisms of heating the dust such as collisions of galaxies 
(\cite{Har87})  and the interaction of the dust with a cluster cooling flow 
(\cite{Les95}) may affect certain galaxies in our sample. It is of interest, in this context, that the far infrared emission in these systems is concentrated in one or both of the interacting components. In the interacting systems that are easily resolved by our measurements, we see no evidence for large amounts of far infrared emission from intranuclear parts. While a large fraction of our sample seems to have undergone collisions, only a few seem to still be so close that their molecular disks could be currently colliding.  The fact that these galaxies continue to produce copious FIR long after their collisions (e.g. $2 \times 10^{7}$ years after the collison for UGC 12915, \cite{Con93}) argues against the collision heating mechanism for most, since the gas cooling time is $\sim$1000 times less than the collision time (\cite{Har87}). NGC 1275 and its associated cooling flow may be special in this regard (Lester, 1995). 

\section{Conclusion}

We have observed the distribution of 100 \micron\ continuum emission in 22 galaxies, most of which have $L_{FIR} > 10^{11} L_{\sun}$. We clearly resolved the emission ($D_{g} > 20$\arcsec, $D_{e} > 12$\arcsec) in 6 of them. We also see some evidence for extension in 7 others. 

For every resolved and possibly resolved source in our sample except for 
MCG+02-04-025, NGC 4151 and NGC 1275, we are able to eliminate all possible methods for FIR production except for starburst heated dust. Of the galaxies we could not resolve, most show correspondingly small starburst regions, and our size limits are consistent with those regions dominating the energetics.

In a few cases, most notably NGC 3110, the disk of the galaxy makes a 
substantial contribution to the FIR flux.  The contribution of the disk of NGC 
3110, and the extended emission outside the bright cores in MCG+02-04-025 and 
NGC 6286 suggest that it is imprudent to assume all of the FIR flux in more 
distant FIR luminous galaxies (that we do not resolve) is concentrated in the 
core.  A better universal model for the FIR flux distribution for these galaxies 
would be a strong central core on top of an extended ``plateau" of emission from 
the disk.

For the systems we resolved, we used the core sizes in a simple, single slab 
emission model to estimate not only the dust temperature T$_{d}$, but also the 
optical depth at 100 \micron\, $\tau_{100}$, assuming a dust emissivity exponent 
of n = 1.5 (emissivity Q $\propto \lambda ^{-n}$).  In addition, when they were 
available, we used fluxes at longer wavelengths that we assumed were still 
associated with the thermal dust emission.  The additional flux measurements 
when combined with the measurements at 60 and 100 \micron\ allow us to solve for 
the emissivity exponent, n, as well.  These measured values of n are all equal 
to or greater than 1.5 and two far exceeded the expected range of 1-2 (e.g. 
\cite{Car92}; see Table 8).

With the values of $\tau_{100}$, we calculated an estimate for the visual 
extinction (via $A_{V} / \tau_{100} \sim 750$ for our galaxy, \cite{Mak85}).  
Assuming that the stars are well mixed with the dust, we can tell how much the 
visible light from the galaxy (and in particular the FIR flux producing region) 
is extinguished.  The single slab model is incapable of dealing with 
geometrical situations that almost surely exist in all of these galaxies such as 
a central condensation of material.  Therefore the A$_{V}$ estimates are only a 
rough indicator of the true extinction of the galaxy.  The $A_{V}$ estimates 
vary from an insignificant value of $\sim$0.2 in NGC 1275 to an extremely high 
value of $\sim$35 in NGC 7469.  Interestingly, and probably not surprisingly, some of the highest extinctions we obtained came from those spiral galaxies which we view edge-on. 

We do not have enough resolving power to determine how the q-value varies within 
these distant FIR luminous galaxies.  We do, however, have enough resolution to 
separate galaxy pairs for which IRAS gives only one flux.  We confirm that UGC 
2369(south), NGC 3110, NGC 6286, UGC 10923(west), NGC 7469, NGC 7541, NGC 7771, 
and UGC 12915 dominate the FIR flux over their companions.  Our observations 
confirm that the q-value for NGC 6286 is unusually low.  We also derive a new q = 2.26 for UGC 12915.  The new value confirms that it is the radio bridge and not low FIR flux from the disk of UGC 12915 that causes the low q-value for the system.

Our project has developed strategies that will be of use for future missions, in particular SOFIA. The large aperture of SOFIA will immediately provide a factor of three improvement in resolution over KAO. New detector arrays deing developed for SOFIA instruments (e.g. HAWC) will better sample this diffraction spot. In addition, the higher sensitivity of SOFIA will not only provide sensitivity to the structure at small scales, but will more specifically make available a large number of asteroids for point source and flux calibration.

\acknowledgments
We would like to thank Beverly Smith, Chris Koresko, and James DiFrancesco for 
help with making the observations, as well as aiding in the data reduction.  We 
are grateful to Cheng-Yue Zhang for writing some of the data reduction software.  
We also wish to thank the staff of the KAO for their support throughout each of 
our flight series. This program was funded under the NASA Airborne Astronomy 
Program NAG2-748.

This research has made use of the NASA/IPAC Extragalactic Database (NED) which 
is operated by the Jet Propulsion Laboratory, Caltech, under contract with the 
National Aeronautics and Space Administration.  The optical light images from 
this paper as well as the data for the Guide Star Catalog (GSC) are from the 
Digital Sky Survey project which used the Palomar Observatory Sky Survey (POSS; 
funded by the National Geographic Society). 

\appendix

\newpage 
\begin{table}
\dummytable\label{table1}

\dummytable\label{table3}

\dummytable\label{table4}

\dummytable\label{table5}

\dummytable\label{table6}

\dummytable\label{table7}

\dummytable\label{table8}

\dummytable\label{table2}
\end{table}

\clearpage

\figcaption[]{The nomenclature and coordinate system of the array is 
schematically shown as it is superimposed on a galaxy. \label{fig1}} 

\figcaption[]{MCG+02-04-025 We show at bottom the shape and average orientation 
of our detector array as it is projected on the sky at the position noted in
Table 1. Detectors 1 and 10 are marked, to facilitate comparison with the plot 
of detected signal. In several cases we make qualitative reference to the 
secondary array (detectors 11-20) in the text, and so these detectors have been 
outlined with dashed rectangles.  The optical image is from the Digital Sky 
Survey (DSS) with north to the top and east to the left.  For scale a bar 
1\arcmin\ long is also shown. The true position angle of the array varied by an 
amount noted in Table 1. The X marks the IRAS position of the galaxy (BGS), 
surrounded by the IRAS error ellipse. At the top of the figure, we show the 
relative flux and errors for each detector in the primary array (squares and 
solid lines). Detector 1 is leftmost, and detector 10 rightmost.  Also included 
are a point source profile scaled and shifted to match the galaxy obserations 
near the peak flux (dashed line), and a point source profile scaled to the IRAS 
flux of the galaxy and shifted to the position where it best fits the entire 
profile (dotted line). \label{fM+02}} Comparison with the dashed line therefore 
tests the extent to which the observed source has a point-like profile, and 
comparison with the dotted line tests how well our observations match a point 
source with the IRAS flux density.

\figcaption[]{III Zw 35 as in Figure~\ref{fM+02}. \label{f3Z35}}
\figcaption[]{UGC 02369 as in Figure~\ref{fM+02}. \label{fU2369}}
\figcaption[]{NGC 1275 as in Figure~\ref{fM+02}. \label{fN1275}}
\figcaption[]{VII Zw 31 as in Figure~\ref{fM+02}. \label{f7Z31}}
\figcaption[]{UGC 05101 as in Figure~\ref{fM+02}. \label{fU5101}}
\figcaption[]{NGC 3110 as in Figure~\ref{fM+02}. \label{fN3110}}
\figcaption[]{NGC 4151 as in Figure~\ref{fM+02}. \label{fN4151}}
\figcaption[]{UGC 08696 as in Figure~\ref{fM+02}. \label{fU8696}}
\figcaption[]{NGC 6090 as in Figure~\ref{fM+02}. \label{fN6090}}
\figcaption[]{NGC 6286 as in Figure~\ref{fM+02}. \label{fN6286}}
\figcaption[]{IRAS 1713+53 as in Figure~\ref{fM+02}. \label{fI1713}}
\figcaption[]{IRAS 1720-00 as in Figure~\ref{fM+02}. \label{fI1720}}
\figcaption[]{UGC 10923 as in Figure~\ref{fM+02}. \label{fU10923}}
\figcaption[]{NGC 7469 as in Figure~\ref{fM+02}. \label{fN7469}}
\figcaption[]{NGC 7541 as in Figure~\ref{fM+02}. \label{fN7541}}
\figcaption[]{Zw 475.056 as in Figure~\ref{fM+02}. \label{fZ475}}
\figcaption[]{NGC 7625 as in Figure~\ref{fM+02}. \label{fN7625}}
\figcaption[]{NGC 7770 as in Figure~\ref{fM+02}. \label{fN7770}}
\figcaption[]{NGC 7771 as in Figure~\ref{fM+02}. \label{fN7771}}
\figcaption[]{Markarian 331 as in Figure~\ref{fM+02}. \label{fMk331}}
\figcaption[]{UGC 12915 as in Figure~\ref{fM+02}. \label{fU12915}}

\end{document}